\documentclass[10pt,journal]{IEEEtran}

\ifCLASSINFOpdf
  \usepackage[pdftex]{graphicx}
  \graphicspath{{./}}
  \DeclareGraphicsExtensions{.pdf,.jpeg,.png}
\else
  \usepackage[dvips]{graphicx}
  \graphicspath{{./fig/}}
  \DeclareGraphicsExtensions{.eps}
\fi
\usepackage{amssymb}
\usepackage[cmex10]{amsmath}

\usepackage{algorithm, algpseudocode}

\usepackage[latin1]{inputenc} 

\newcommand{\ve}{\mathbf}
\newcommand{\m}{\mathbf}

\newcommand{\mf}[1]{\mathbf{\tilde{\mathbf{#1}}}} 
\newcommand{\vef}[1]{\mathbf{\tilde{\mathbf{#1}}}} 

\begin{document}
%
\title{Non-Systematic Complex Number RS Coded OFDM by Unique Word Prefix}
%

\author{Mario~Huemer,~\IEEEmembership{Senior Member,~IEEE}, Christian
  Hofbauer,~\IEEEmembership{Student Member,~IEEE}\\ and Johannes Huber,~\IEEEmembership{Fellow,~IEEE}
\thanks{%
Christian Hofbauer was funded by the European Regional Development Fund and the Carinthian Economic Promotion Fund (KWF) under grant 20214/15935/23108. He is currently funded by the Austrian Science Fund (FWF): I683-N13.

The authors are with the Institute of Networked an Embedded Systems,
Alpen-Adria-Universit\"at Klagenfurt, Austria  (e-mail:
mario.huemer@uni-klu.ac.at; chris.hofbauer@uni-klu.ac.at), and with the
Institute for Information Transmission, University of Erlangen-Nuremberg, Germany
(e-mail: huber@lnt.de), respectively.

Copyright (c) 2011 IEEE. Personal use of this material is permitted. Permission
from IEEE must be obtained for all other uses, in any current or future media,
including reprinting/republishing this material for advertising or promotional
purposes, creating new collective works, for resale or redistribution to
servers or lists, or reuse of any copyrighted component of this work in other
works.

Digital Object Identifier 10.1109/TSP.2011.2168522
}}

\maketitle

\begin{abstract}
In this paper we expand our recently introduced concept of UW-OFDM (unique word
orthogonal frequency division multiplexing). In UW-OFDM the cyclic prefixes
(CPs) are replaced by deterministic sequences, the so-called unique words
(UWs). The UWs are generated by appropriately loading a set of redundant
subcarriers. By that a systematic complex number Reed Solomon (RS) code
construction is introduced in a quite natural way, because an RS code may be
defined as the set of vectors, for which a block of successive zeros occurs in
the other domain w.r.t. a discrete Fourier transform. (For a fixed block
different to zero, i.e., a UW, a coset code of an RS code is generated.)
A remaining problem in the original systematic coded UW-OFDM concept is
the fact that the redundant subcarrier symbols disproportionately contribute to
the mean OFDM symbol energy. In this paper we introduce the
concept of non-systematic coded UW-OFDM, where the redundancy is no longer
allocated to dedicated subcarriers, but distributed over all subcarriers. We
derive optimum complex valued code generator matrices matched to the BLUE (best
linear unbiased estimator) and to the LMMSE (linear minimum mean square error)
data estimator, respectively. With the help of simulations we highlight the
advantageous spectral properties and the superior BER (bit error ratio)
performance of non-systematic coded UW-OFDM compared to systematic coded
UW-OFDM as well as to CP-OFDM in AWGN (additive white Gaussian noise) and in frequency
selective environments. 
\end{abstract}


\IEEEpeerreviewmaketitle

\section{Introduction}

\begin{figure}[!ht]
\centering
\includegraphics[width=3.3in]{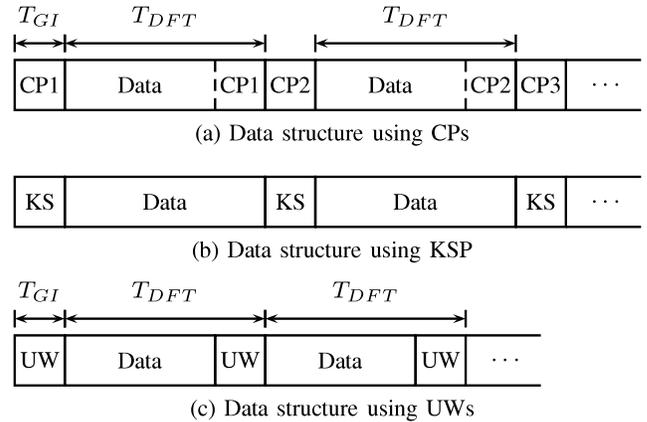}
\caption{Transmit data structure using a CP (a), a KS (b) or a UW (c).}
\label{fig:sym_structures_UW}
\end{figure}

IN \cite{Huemer10_1}, \cite{Onic10_1} we introduced an OFDM (orthogonal frequency division multiplexing) signaling scheme, where the usual cyclic prefixes (CPs) \cite{VanNee00} are replaced by deterministic sequences, that we call unique words (UWs). A related but -- when regarded in detail -- also very different scheme is KSP (known symbol padded)-OFDM  \cite{Tang07}, \cite{Welden08}. Fig.~\ref{fig:sym_structures_UW}a -- \ref{fig:sym_structures_UW}c compare the CP-, KSP-, and UW-based OFDM transmit data structures.
In CP- as well as in UW-OFDM the linear convolution of the transmit signal with
the channel impulse response is transformed into a cyclic convolution such that
the discrete Fourier transform (DFT) can diagonalize the channel in the
frequency domain. However,
different to the CP, the UW is part of the DFT-interval as indicated in
Fig.~\ref{fig:sym_structures_UW}. Furthermore, the CP is a random sequence,
whereas the UW is deterministic. Hence, the UW can optimally be designed for
particular needs like synchronization and/or channel estimation purposes at the
receiver side. The broadly known KSP-OFDM uses a structure similar to UW-OFDM,
since the known symbol (KS) sequence is deterministic as well. The most
important difference between KSP- and UW-OFDM is the fact, that the UW is part
of the DFT interval, whereas the KS is not. The generation of the UW within the
DFT-interval introduces correlations among the subcarriers, which can
advantageously be exploited by the receiver to improve the BER (bit error
ratio) performance. Whilst in both schemes the deterministic sequences can be
used for synchronization and channel estimation purposes, these correlations
are not present in KSP-OFDM. We notice that KSP-OFDM coincides with ZP-OFDM (zero padded OFDM) \cite{Lin11}, if the KS sequence is set to zero. \smallskip

Since UW-OFDM time domain symbols contain a block of fixed samples, i.e., the UW, the set of all corresponding vectors in discrete frequency domain forms a coset to a Reed Solomon code (RS code). Usually RS codes of length $n$ are defined for a finite field $\mathbb{F}_Q$ using an element $w\in\mathbb{F}_Q$ of order $n$, $n\cdot l=Q-1$, with $n,l,Q\in\mathbb{N}$ to define a discrete Fourier transform $\mathbb{F}_Q^n\rightarrow\mathbb{F}_Q^n$ in $\mathbb{F}_Q$. The set of codewords is specified by the fact, that the (inverse) DFT of all codewords contains a block of $d_{min}-1$ successive zeros, where $d_{min}$ is the minimum Hamming distance of the RS code. If this block of $d_{min}-1$ successive symbols differs from zero, but is also fixed for all codewords, a coset code to an RS code is generated in the other domain w.r.t. this Fourier transform with the same minimum distance $d_{min}$, c.f. \cite{Blahut03}.

All these definitions apply for the field of complex numbers and a usual DFT of
length $n$ as well. Thus, in UW-OFDM the set of frequency domain data vectors
defines a coset code to an RS code in a quite natural way. In contrast to the usual
approach using RS codes over a finite field, e.g. $\mathbb{F}_{2^8}$, for an
outer code in a concatenated code scheme, in UW-OFDM we have an inner RS code
over the field of complex numbers if any further channel coding scheme is
applied, i.e., the OFDM guard space is here additionally exploited for
redundancy of an inner channel coding scheme in a natural way. In
\cite{Hofbauer10_03} we showed that algebraic decoding of the introduced
complex number RS code leads to solving an ill-conditioned system of equations
which is extremely sensitive to noise. It turns out that the application of data
estimation approaches like the BLUE (best linear unbiased estimator) or the
LMMSE (linear minimum mean square error) estimator, cf. \cite{Huemer11_1}, is much more appropriate than algebraic decoding. \smallskip

For SC/FDE (single carrier/frequency domain equalization) systems \cite{Sari94a}--\cite{Reinhardt05}, the benefits of UW based transmission have already sufficiently been studied \cite{Imec00}--\cite{Witschnig02b}, \cite{Huemer03a}--\cite{Witschnig03a}. The introduction of UWs in SC/FDE systems is straightforward, since the data symbols as well as the UW symbols are defined in time domain. In UW-OFDM the data symbols are defined in frequency domain, whereas the UW symbols are defined in time domain, which leads to some difficulties. In \cite{Huemer10_2} we discussed the similarities and differences of the UW approach for OFDM and SC/FDE. \smallskip

In our concept described in \cite{Huemer10_1}, \cite{Huemer11_1} we suggested to generate UW-OFDM symbols by appropriately loading so-called redundant subcarriers. The minimization of the energy contribution of the redundant subcarriers turned out to be a challenge. We solved the problem by generating a zero UW in a first step, and by adding the desired UW in a separate second step. We showed that this approach generates OFDM symbols with much less redundant energy \cite{Onic10_1} than a single step or direct UW generation approach as e.g., described in \cite{Cendrillon01}. Additionally, we optimized the positions of the redundant subcarriers to further reduce their energy contribution. 
Several other attempts of applying UWs in OFDM systems can be found in the literature, e.g. in \cite{Muck06}-\cite{Jingyi02}. In all those approaches the guard interval and thus the UW is not part of the DFT-interval. Therefore, and in contrast to our UW-OFDM concept no coding is introduced by these schemes.  \smallskip

Our systematic complex number RS coded UW-OFDM concept presented in
\cite{Huemer10_1} and shortly reviewed in Sec. \ref{sec:uw} of the present
paper still suffers from a disproportionately high energy contribution of the
redundant subcarriers. In \cite{Hofbauer10_2} we tackled this problem by increasing the number of
redundant subcarriers while keeping the length of the UW constant. On the one
hand this approach in fact leads to a reduction of the redundant energy
contribution and to an improved BER performance, but on the other hand the
bandwidth efficiency decreases compared to the original concept. In
\cite{Huemer10_3} we introduced another approach that also focuses on the
redundant energy contribution. Here we achieved the reduction of the redundant
energy by allowing some systematic noise within the guard interval. This method
clearly outperforms the original UW-OFDM approach, however, a remaining penalty
is the fact that the UW is disturbed to some extent. In the present paper we introduce
a different and much more favorable approach to overcome the shortcomings of
the original UW-OFDM concept. We no longer primarily focus on the redundant
energy reduction. Instead, we suggest to distribute the redundant energy over all
subcarriers, and we define cost functions that take the overall transceiver
performance (including the data estimation) into account. The corresponding
UW-OFDM symbol generation procedure introduces a \emph{non}-systematic complex
number RS code construction (cf. Sec. \ref{sec:syst_coding}) which can be
described by appropriate code generator matrices. For the data estimation we
apply two different approaches, namely the BLUE and the LMMSE
estimator. Sec. \ref{sec:optimization} is dedicated to the solutions of the arising
optimization problems. At first we solve the optimization problems numerically,
thereafter we analytically derive a number of highly interesting general
properties of optimum code generator matrices and the implications for the
overall system approach. Moreover, we discuss the properties of two particular
numerically found code generator matrices. In Sec. \ref{sec:scfde} we show that
non-systematic coded UW-OFDM can be converted into a UW-SC/FDE system by
choosing a specific constructed optimum code generator matrix. Finally,
simulation results are presented in Sec. \ref{sec:simulations}. We compare the
novel UW-OFDM approach against our original systematic coded UW-OFDM concept
and against a classical CP-OFDM system, as a reference system we use the IEEE
802.11a WLAN (wireless local area network) standard. The spectral advantages
are discussed, and BER simulation results  are presented for the AWGN channel as well
as for frequency selective indoor scenarios. For the latter case we
additionally investigate the impact of channel estimation errors on the BER
performance. The results highlight the advantageous properties of the proposed scheme. \bigskip

\noindent
\textit{Notation:} Lower-case bold face variables ($\ve{a},\ve{b}$,...) indicate vectors, and upper-case bold face variables 
($\m{A},\m{B}$,...) indicate matrices. To distinguish between time and frequency domain variables, we use a tilde
to express frequency domain vectors and matrices ($\vef{a},\mf{A}$,...), respectively. We further use $\mathbb{R}$ to denote the set 
of real numbers, $\mathbb{C}$ to denote the set of complex numbers, $\m{I}$ to
denote the identity matrix, $(\cdot)^T$ to denote transposition, $(\cdot)^*$ to
denote complex conjugation, $(\cdot)^H$ to denote conjugate transposition, $E[\cdot]$ to denote expectation, and $\mathrm{tr}\{\cdot\}$ to denote the trace operator. For all signals and systems the usual equivalent complex baseband representation is applied. \bigskip

\section{Review of Systematic Coded UW-OFDM} \label{sec:uw}
\subsection{Unique Word Generation}
We briefly review our original approach of introducing unique words in OFDM
time domain symbols, for further details see \cite{Huemer10_1},
\cite{Onic10_1}. Let $\ve{x}_u\in\mathbb{C}^{N_u \times 1}$ be a predefined
sequence which we call unique word. This unique word shall form the tail of
each OFDM time domain symbol vector. Hence, an OFDM time domain symbol vector,
as the result of a length-$N$-IDFT (inverse DFT), consists of two parts and is
of the form $\begin{bmatrix}\ve{x}_d^T & \ve{x}_u^T \end{bmatrix}^T\in
\mathbb{C}^{N \times 1}$, whereat only $\ve{x}_d\in\mathbb{C}^{(N-N_u) \times
  1}$ is random and affected by the data. In the concept suggested in
\cite{Huemer10_1}, \cite{Onic10_1} we generate an OFDM symbol $\ve{x}
= \begin{bmatrix}\ve{x}_d^T & \ve{0}^T\end{bmatrix}^T$ with a zero UW in a
  first step, and we determine the final transmit symbol $\ve{x}' = \ve{x}
  + \begin{bmatrix}\ve{0}^T & \ve{x}_u^T\end{bmatrix}^T$ by adding the desired
    UW in time domain in a second step. As in conventional OFDM, the QAM data
    symbols (denoted by the vector $\vef{d}\in\mathbb{C}^{N_d \times 1}$) and
    the zero subcarriers (at the band edges and at DC) are specified as part of
    the frequency domain vector $\vef{x}$, but here in addition the zero word
    is specified in time domain as part of the vector
    $\ve{x}=\m{F}_N^{-1}\vef{x}$. $\m{F}_N$ denotes the length-$N$-DFT
    matrix with elements
    $[\m{F}_N]_{kl}=\mathrm{e}^{-\mathrm{j}\frac{2\pi}{N}kl}$ for
    $k,l=0,1,...,N-1$. The system of equations $\ve{x}=\m{F}_N^{-1}\vef{x}$
    with the introduced features can, e.g., be fulfilled by spending a set of redundant subcarriers. We let the redundant subcarrier symbols form the vector $\vef{r}\in\mathbb{C}^{N_r \times 1}$ with $N_r=N_u$, we further introduce a permutation matrix $\m{P}\in\mathbb{C}^{(N_d+N_r) \times (N_d+N_r)}$, and form an OFDM symbol (containing $N-N_d-N_r$ zero subcarriers) in frequency domain by 
\begin{equation}
	\vef{x} = \m{B} \m{P} \begin{bmatrix} \vef{d} \\ \vef{r} \end{bmatrix}. \label{equ:2}
\end{equation}
$\m{B}\in\mathbb{C}^{N \times (N_d+N_r)}$ inserts the usual zero subcarriers. It consists of zero-rows at the positions of the zero subcarriers, and of appropriate unit row vectors at the positions of data subcarriers. We will detail the reason for the introduction of the permutation matrix $\m{P}$ and its specific construction shortly below. 
The time - frequency relation of the OFDM symbol (before adding the desired UW)
can now be written as
\begin{equation}
	\m{F}_N^{-1} \m{B} \m{P} \begin{bmatrix} \vef{d} \\ \vef{r} \end{bmatrix} = \begin{bmatrix}\ve{x}_d \\ \ve{0} \end{bmatrix}.
	\label{equ:6}
\end{equation}
With 
\begin{equation}
		\m{M}=\m{F}_N^{-1} \m{B} \m{P}= \begin{bmatrix} \m{M}_{11} & \m{M}_{12} \\ \m{M}_{21} & \m{M}_{22}\end{bmatrix}, \label{equ:syst019}
\end{equation}
where $\m{M}_{ij}$ are appropriate sized sub-matrices, it follows that $\m{M}_{21} \vef{d} + \m{M}_{22} \vef{r} = \ve{0}$, and hence 
$\vef{r} = -\m{M}_{22}^{-1}\m{M}_{21} \vef{d}$. With the matrix 
\begin{equation}
	\m{T} = -\m{M}_{22}^{-1}\m{M}_{21} \in\mathbb{C}^{N_r \times N_d}, \label{equ:syst020}
\end{equation}
the vector of redundant subcarrier symbols can thus be determined by the linear mapping
\begin{equation}
\vef{r} = \m{T} \vef{d}. \label{equ:3}
\end{equation}
Equation \eqref{equ:3} introduces correlations in the vector $\vef{x}$ of frequency domain samples of an OFDM symbol. The construction of $\m{T}$, and thus also the energy of the redundant subcarrier symbols, depends on the choice of $\m{P}$. The mean symbol energy $E_{\ve{x}'}=E[(\ve{x}')^H\ve{x}']$ can be calculated to  
\begin{equation}
		E_{\ve{x}'} = 	\frac{1}{N} (\underbrace{N_d \sigma_d^2}_{E_{\vef{d}}}  + \underbrace{\sigma_d^2 																								\mathrm{tr}(\m{T}\m{T}^H)}_{E_{\vef{r}}}) + \underbrace{\ve{x}_u^H\ve{x}_u}_{E_{\ve{x}_u}}, 																		\label{equ:trans019}
\end{equation}
cf. \cite{Onic10_1}. $E_{\vef{d}}/N$ and $E_{\vef{r}}/N$ describe the
contributions of the data and the redundant subcarrier symbols to the total
mean symbol energy before the addition of the desired UW, respectively, and
$E_{\ve{x}_u}$ describes the contribution of the UW. It turns out that the
energy contribution $E_{\vef{r}}/N$ of the redundant subcarrier symbols almost
explodes without the use of an appropriate permutation matrix, or equivalently for $\m{P}=\m{I}$. In \cite{Huemer10_1} we therefore suggested to choose $\m{P}$ by minimizing the symbol energy $E_{\ve{x}'}$ or equivalently by minimizing the energy-based cost function
\begin{equation}
  J_E = \frac{\sigma_d^2}{N}\mathrm{tr}\left\{\m{T}\m{T}^H\right\}. \label{equ:uw001}
\end{equation}
Note that $\m{T}$ is derived from \eqref{equ:syst019} and \eqref{equ:syst020}.
\smallskip

\noindent 
\textit{Example 1:} For the parameter choice $N=64$, $N_u=16$, and an index set of the zero subcarriers given by \{0, 27, 28,...,37\}  (these parameters are taken from the IEEE 802.11a WLAN standard \cite{IEEE99}, see also Table \ref{tab2}), we have $N_r = 16$ and $N_d = 36$. The optimum index set for the redundant subcarriers as a result of minimizing the cost function in \eqref{equ:uw001} is \{2, 6, 10, 14, 17, 21, 24, 26, 38, 40, 43, 47, 50, 54, 58, 62\}; cf. \cite{Huemer10_1}. This choice can easily also be described by \eqref{equ:2} with appropriately constructed matrices $\m{B}$ and $\m{P}$. We assume uncorrelated and zero mean QAM data symbols with the covariance matrix $\m{C}_{\tilde{d}\tilde{d}}= \sigma_d^2\m{I}$. 
\begin{figure}[!ht]
\centering
\includegraphics[width=3.5in]{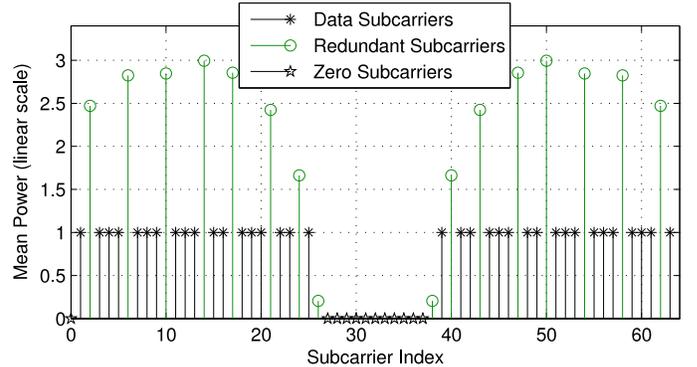}
\caption{Mean power of individual subcarrier symbols for Example 1.}
\label{fig:mean_power}
\end{figure}
Fig. \ref{fig:mean_power} shows the mean power values of all individual subcarrier symbols for the chosen parameter setup for the case the UW is the zero word $\ve{x}_u=\ve{0}$ and for $\sigma_d^2=1$. The optimized mean power values of the redundant subcarrier symbols are the elements of the vector $\sigma_d^2 \mathrm{diag}\left(\m{T}\m{T}^H\right)$ evaluated for the optimum permutation matrix $\m{P}$.

\subsection{Interpretation as a Systematic Complex Valued Reed-Solomon Code}
\begin{figure}[!t]
\centering
\includegraphics[width=2.8in]{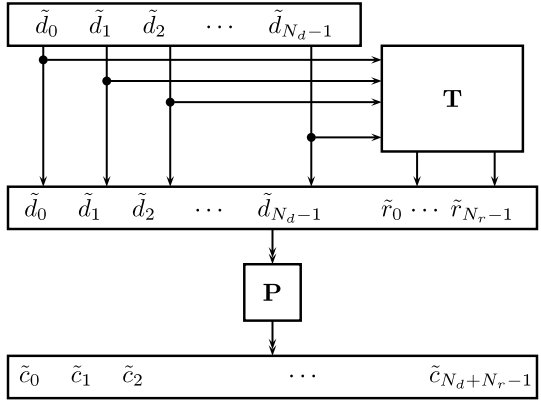}
\caption{\footnotesize{Codeword generator for the systematic code described by $\m{G}$.}}
\label{fig:codeword_generator}
\end{figure}

With 
\begin{equation}
	\m{G} = \m{P} \begin{bmatrix} \m{I} \\ \m{T} \end{bmatrix} \in\mathbb{C}^{(N_d+N_r) \times N_d} \label{equ:trans023}
\end{equation}
we can interpret
\begin{equation}
	\vef{c} = \m{P} \begin{bmatrix} \vef{d} \\ \vef{r} \end{bmatrix}
					= \m{P} \begin{bmatrix} \m{I} \\ \m{T} \end{bmatrix} \vef{d}
					= \m{G} \vef{d} \label{equ:trans024}
\end{equation}
($\vef{c}\in\mathbb{C}^{(N_d+N_r) \times 1}$) as a codeword of a
\emph{systematic} complex number Reed Solomon code construction with the code
generator matrix $\m{G}$. As
already mentioned above an RS code with minimum Hamming distance $d_{min}$ may be defined as the set of codewords, which all show a block of $d_{min}-1$ consecutive zeros in their spectral transform w.r.t. a Fourier transform defined in the (elsewhere usually finite) field from which the code symbols are taken; c.f. \cite{Blahut03}. Here, simply time and frequency domains are interchanged and the field is the set of complex numbers. Fig. \ref{fig:codeword_generator} graphically illustrates the generation of a codeword $\vef{c}=[\tilde{c}_0,\tilde{c}_1,...,\tilde{c}_{N_d+N_r-1}]^T$. 
Using \eqref{equ:trans024}, and with the frequency domain version of the UW $\vef{x}_u=\m{F}_N \begin{bmatrix} \ve{0}^T&\ve{x}_u^T \end{bmatrix}^T$ the transmit symbol can now also be written as
\begin{equation}
	\ve{x}' = \m{F}_N^{-1} (\m{B}\m{G}\vef{d}+\vef{x}_u). \label{eq:tx_generation2}
\end{equation}

\subsection{System Model and Preparatory Steps}
After the transmission over a dispersive (e.g., multipath) channel a received OFDM time domain symbol can be modeled as
\begin{align}
	\ve{y}_r 	&= \m{H}_c \ve{x}'+ \ve{n} \\
						&= \m{H}_c \m{F}_N^{-1} (\m{BG}\vef{d}+\vef{x}_u) + \ve{n},
\end{align}
cf. \eqref{eq:tx_generation2}, where $\ve{n}\in\mathbb{C}^{N \times 1}$ represents a zero-mean Gaussian (time domain) noise vector with the covariance matrix $\sigma_n^2 \m{I}$, and $\m{H}_c\in\mathbb{C}^{N\times N}$ denotes a cyclic convolution matrix originating from the zero-padded vector of channel impulse response coefficients $\ve{h}_c\in\mathbb{C}^{N\times 1}$. After applying a DFT to obtain $\vef{y}_r=\m{F}_N\ve{y}_r$, we exclude the zero subcarriers from further operation, which leads to the down-sized vector $\vef{y}_d=\m{B}^T\vef{y}_r$ with $\vef{y}_d\in\mathbb{C}^{(N_d+N_r) \times 1}$:
\begin{equation}
		\vef{y}_d = \m{B}^T \m{F}_N \m{H}_c \m{F}_N^{-1} (\m{BG}\vef{d} + \vef{x}_u) + \m{B}^T \m{F}_N \ve{n}.
\end{equation}
The matrix $\mf{H}_c=\m{F}_N \m{H}_c \m{F}_N^{-1}$ is diagonal and contains the sampled channel frequency response on its main diagonal. $\mf{H} = \m{B}^T \m{F}_N \m{H}_c \m{F}_N^{-1} \m{B}$ with $\mf{H}\in\mathbb{C}^{(N_d+N_r) \times (N_d+N_r)}$ is a down-sized version of the latter excluding the entries corresponding to the zero subcarriers. The received symbol can now be written in the form of the \emph{affine} model
\begin{equation}
		\vef{y}_d = \mf{H}\m{G}\vef{d} + \mf{H}\m{B}^T \vef{x}_u + \m{B}^T \m{F}_N \ve{n}. \label{equ:system_model_002}
\end{equation}
Note that (assuming that the channel matrix $\mf{H}$ or at least an estimate of the same is available) $\mf{H} \m{B}^T \vef{x}_u$ represents the known portion contained in the received vector $\vef{y}_d$ originating from the UW. As a preparatory step to the data estimation procedure the UW influence is subtracted to obtain the corrected symbol $\vef{y}=\vef{y}_d - \mf{H} \m{B}^T \vef{x}_u$ in the form of the \emph{linear} model
\begin{equation}
		\vef{y} = \mf{H}\m{G}\vef{d} + \vef{v}, \label{equ:system_model_001}
\end{equation}
with the noise vector $\vef{v} = \m{B}^T \m{F}_N \ve{n}$. The vector $\vef{y}$ serves as the input for the data estimation (or equalization) procedure. In the following we will consider linear data estimators of the form
\begin{equation}
	\widehat{\vef{d}} = \m{E} \vef{y}, \label{equ:001}
\end{equation}
where $\m{E}\in\mathbb{C}^{N_d \times (N_d+N_r)}$ describes the equalizer. 

\subsection{Optimum Linear Data Estimators}
One way to look for an optimum data estimator is to assume the data vector to be deterministic but unknown, and to search for unbiased estimators. In order for the estimator to be unbiased we require
\begin{equation}
	E[\widehat{\vef{d}}] = E[\m{E} \vef{y}] = \m{E}E[\mf{H}\m{G}\vef{d} + \vef{v}] = \m{E}\mf{H}\m{G}\vef{d} = \vef{d}. 
\end{equation}
Consequently, the unbiased constraint takes on the form
\begin{equation}
	\m{E}\mf{H}\m{G} = \m{I}. \label{equ:ZF_006}
\end{equation}
Equ. \eqref{equ:ZF_006} is equivalent to the ZF (zero forcing) criterion for linear equalizers. The optimum solution which is commonly known as the best linear unbiased estimator, and which is equivalent to the optimum linear ZF equalizer, is found by applying the Gauss-Markov theorem, cf.~\cite{Kay93}, to the linear model in \eqref{equ:system_model_001}. The solution is given by
\begin{equation}
	\m{E}_\mathrm{BLUE} = (\m{G}^H\mf{H}^H \m{C}_{\tilde{v}\tilde{v}}^{-1} \mf{H}\m{G})^{-1} \m{G}^H\mf{H}^H 																									\m{C}_{\tilde{v}\tilde{v}}^{-1}. \label{equ:ZF_007}
\end{equation}
We note that since the noise in \eqref{equ:system_model_001} is assumed to be Gaussian, \eqref{equ:ZF_007} is also the MVU (minimum variance unbiased) estimator. With the noise covariance matrix $\m{C}_{\tilde{v}\tilde{v}} = E \left[\vef{v} \vef{v}^H \right]=N \sigma_n^2 \m{I}$ we obtain 
\begin{equation}
	\m{E}_\mathrm{BLUE} = (\m{G}^H\mf{H}^H \mf{H}\m{G})^{-1} \m{G}^H\mf{H}^H. \label{equ:ZF_008}
\end{equation}
The covariance matrix of $\widehat{\vef{d}}=\m{E}_\mathrm{BLUE}\vef{y}$, or equivalently, the covariance matrix of the error $\vef{e} = \vef{d}-\widehat{\vef{d}}$ immediately follows to
\begin{equation}
	\m{C}_{\tilde{e}\tilde{e}} = N \sigma_n^2 (\m{G}^H\mf{H}^H \mf{H}\m{G})^{-1}. \label{equ:ZF_009}
\end{equation}
The most common linear data estimator is the LMMSE estimator which belongs to the class of the Bayesian estimators. In the Baysian approach the data vector is assumed to be the realization of a random vector instead of being deterministic but unknown as assumed above. By applying the Bayesian Gauss-Markov theorem \cite{Kay93} to \eqref{equ:system_model_001}, where we now assume $\vef{d}$ to be the realization of a random vector, the LMMSE equalizer follows to
\begin{equation}
	\m{E}_\mathrm{LMMSE} = (\m{G}^H\mf{H}^H \mf{H}\m{G} + \frac{N \sigma_n^2}{\sigma_d^2}\m{I})^{-1} \m{G}^H\mf{H}^H.       			              \label{equ:LMMSE_005}
\end{equation}
Expression \eqref{equ:LMMSE_005} shows huge similarity to the BLUE in \eqref{equ:ZF_008}. For $\sigma_n^2 = 0$ the LMMSE equalizer and the BLUE are identical. The covariance matrix of the error $\vef{e} = \vef{d}-\widehat{\vef{d}}$ is given by
\begin{equation}
	\m{C}_{\tilde{e}\tilde{e}} = N \sigma_n^2 (\m{G}^H\mf{H}^H \mf{H}\m{G} + \frac{N \sigma_n^2}{\sigma_d^2}\m{I})^{-1}.       			              \label{equ:LMMSE_006}
\end{equation}

\section{UW Generation by Optimum Non-Systematic Coding} \label{sec:syst_coding}
In Sec. \ref{sec:uw} we chose the positions of the redundant subcarriers (represented by the choice of the permutation matrix $\m{P}$) such that the
redundant energy becomes minimum. For that we had to minimize the cost function $J_\text{E}$ in \eqref{equ:uw001}. Nevertheless, the mean power of the redundant subcarrier symbols is still considerably higher than that of the data symbols; cf. Fig. \ref{fig:mean_power}. In this section we present a novel and completely different approach to optimize the overall system performance by adapting our original concept as follows: 
\begin{enumerate}
\item We give up the idea of dedicated redundant subcarriers, and we allow to spread the redundant energy over all codeword symbols.
\item We define new cost functions that additionally take the receiver processing into account. Instead of purely focusing on the redundant energy, we define performance measures based on the sum of the error variances at the output of the data estimator.
\end{enumerate}
It will turn out that this approach significantly outperforms CP-OFDM and also our original systematic coded UW-OFDM.  

\subsection{The Idea of Non-Systematic Coding in UW-OFDM}
With the introduction and optimization of the permutation matrix $\m{P}$ we minimized the energy contribution of the redundant subcarrier symbols. From the optimum choice of the permutation matrix $\m{P}$ we learned that the redundant subcarrier symbols shall be distributed approximately equidistantly over the codeword $\vef{c}$; cf. Fig \ref{fig:mean_power}. This means that the redundant energy is not concentrated in bundles of subcarrier symbols, but it is spread out over the codeword. Nevertheless, the portions of the redundant energy are only concentrated on the dedicated redundant subcarrier symbol positions. However, from this equidistant distribution of the redundant energy one could guess, that it might make sense to distribute the redundancy over all subcarrier symbols. If we do so we can no longer speak of dedicated redundant subcarriers, since every subcarrier will then carry an amount of redundant energy instead. We incorporate this idea into our UW-OFDM symbol generation process by replacing $\m{G}$ as defined in \eqref{equ:trans023} 
by a code generator matrix $\breve{\m{G}}$ (of the same size as $\m{G}$) which spreads the redundancy over all codeword symbols. The code described by $\breve{\m{G}}$ can then be interpreted as a non-systematic code since the original data symbols $\vef{d}$ will not appear in the codeword
\begin{equation}
	\vef{c} = \breve{\m{G}}\vef{d} 
\end{equation}
any longer. $\breve{\m{G}}$ distributes portions of a single data symbol over
all (or at least several) codeword symbols, and it additionally adds
redundancy. Consequently, and analogical to $\m{G}$, $\breve{\m{G}}$ can be
interpreted as a mixture of a linear dispersive preprocessor (or
channel-independent precoder, cf. \cite{Lin11},) and a channel coder. However, $\breve{\m{G}}$ significantly differs from $\m{G}$
in the specific way how data and redundancy are spread over the codeword. \smallskip

In the following we will formulate optimization criteria from which $\breve{\m{G}}$ shall be derived. Following the way we optimized the permutation matrix $\m{P}$, we could again think of a redundant energy minimization. However, since the redundant energy will now be smeared over all subcarrier symbols it is not clear how to enforce this. Therefore, we no longer primarily focus on the redundant energy reduction, but we aim for optimization criteria that take the complete transceiver processing into account.

\subsection{Transceiver Cost Function for the BLUE}
Clearly the linear data estimators in \eqref{equ:ZF_008} and \eqref{equ:LMMSE_005} can also be used for non-systematic coded UW-OFDM, we only have to substitute $\m{G}$ by $\breve{\m{G}}$. We first focus on the BLUE given by \eqref{equ:ZF_008}. A possible approach to optimize the overall transceiver performance is to choose the code generator matrix $\breve{\m{G}}$ such that the sum over the error variances after the data estimation becomes minimum. With \eqref{equ:ZF_009} this would lead to the cost function
\begin{equation}
	J = \mathrm{tr}\{\m{C}_{\tilde{e}\tilde{e}}\} 
						= N\sigma_n^2\mathrm{tr}\left\{(\breve{\m{G}}^H\mf{H}^H\mf{H}\breve{\m{G}})^{-1}\right\}. \label{equ:syst003}
\end{equation}
We are aiming for a code generator matrix design which shall be done only once during system design. Because of that reason the dependence of the cost function on the particular channel $\mf{H}$ is inappropriate. We therefore suggest to look for an optimum $\breve{\m{G}}$ for the case $\mf{H}=\m{I}$, that is the AWGN channel case. $J$ then reduces to
\begin{equation}
	J = N\sigma_n^2\mathrm{tr}\left\{(\breve{\m{G}}^H\breve{\m{G}})^{-1}\right\}.  \label{equ:syst006}
\end{equation}
In the simulation section we will demonstrate that the finally derived
non-systematic coded UW-OFDM systems not only perform superior in the AWGN
channel case, but also and particularly in frequency selective
channels\footnote{It will turn out, that the solution to the formulated
  optimization problem is not unambiguous.}. As we will see this comes from the
advantageous combination of the channel coding and dispersive preprocessing abilities of the
optimized code generator matrices. But let us come back to the formulation of
an appropriate optimization criterion and to $J$ as given in
\eqref{equ:syst006}: We could now try to minimize $J$ for a given $\sigma_n^2$,
where the particular choice of $\sigma_n^2$ is obviously irrelevant. However, different choices of $\breve{\m{G}}$ lead to different mean OFDM symbol energies and consequently to different ratios $E_s/\sigma_n^2$, where $E_s$ denotes the mean energy of an individual QAM data symbol. Since it is not desirable to reach the goal of a small sum of error variances at the cost of a huge transmit energy, it is much more reasonable and fair to fix the ratio $c=E_s/\sigma_n^2$ during the optimization. To obtain an expression for $E_s$ we calculate the mean OFDM symbol energy $E_\ve{x}$ (for the case of a zero UW) first:
\begin{eqnarray}
	E_\ve{x} &=& E[\ve{x}^H\ve{x}] \nonumber \\
			&=& E[(\m{F}_N^{-1}\m{B}\breve{\m{G}}\vef{d})^H \m{F}_N^{-1}\m{B}\breve{\m{G}}\vef{d}] \nonumber \\
			&=& \frac{1}{N} E[\vef{d}^H\breve{\m{G}}^H\m{B}^T\m{F}_N\m{F}_N^{-1}\m{B}\breve{\m{G}}\vef{d}] \nonumber \\
			&=& \frac{1}{N} E[\vef{d}^H\breve{\m{G}}^H\breve{\m{G}}\vef{d}]
\end{eqnarray}
With $\ve{a}^H\ve{b}=\mathrm{tr}\{\ve{b}\ve{a}^H\}$ we can further write
\begin{eqnarray}
	E_\ve{x} &=& \frac{1}{N} E[\mathrm{tr}\{\breve{\m{G}}^H\breve{\m{G}}\vef{d}\vef{d}^H\}] \nonumber \\
			&=& \frac{1}{N} \mathrm{tr}\{E[\breve{\m{G}}^H\breve{\m{G}}\vef{d}\vef{d}^H]\} \nonumber \\
			&=& \frac{\sigma_d^2}{N}\mathrm{tr}\{\breve{\m{G}}^H\breve{\m{G}}\}. \label{equ:syst041}
\end{eqnarray}
The mean QAM data symbol energy $E_s$ follows to $E_s = E_\ve{x}/N_d$. With $c=E_s/\sigma_n^2$ we obtain 
\begin{equation}
	\sigma_n^2=\frac{E_s}{c} = \frac{\sigma_d^2\mathrm{tr}\{\breve{\m{G}}^H\breve{\m{G}}\}}{c N N_d}.  \label{equ:syst004} 
\end{equation}
Inserting \eqref{equ:syst004} into \eqref{equ:syst006} finally yields the cost function
\begin{equation}
	J_\text{BLUE} = \frac{\sigma_d^2}{c N_d} \mathrm{tr}\left\{\breve{\m{G}}^H\breve{\m{G}}\right\}
											\mathrm{tr}\left\{(\breve{\m{G}}^H\breve{\m{G}})^{-1}\right\}.  \label{equ:syst007}
\end{equation}
The cost function $J_\text{BLUE}$ measures the overall transceiver performance at a fixed ratio $E_s/\sigma_n^2$. 
However, the particular choice of the ratio $c=E_s/\sigma_n^2$ is obviously irrelevant for the searching of optimum code generator matrices. \smallskip

An optimum code generator matrix $\breve{\m{G}}$ shall consequently be found by minimizing $J_\text{BLUE}$, but in addition, in order that the zero UW is generated in the time domain, $\breve{\m{G}}$ has to be constrained to fulfill
\begin{equation}
	\m{F}_N^{-1}\m{B}\breve{\m{G}}\vef{d} = \begin{bmatrix} \ve{x}_d \\ \ve{0} \end{bmatrix} \label{equ:syst016}
\end{equation}
for every possible data vector $\vef{d}$, or equivalently
\begin{equation}
	\m{F}_N^{-1}\m{B}\breve{\m{G}} = \begin{bmatrix} \m{*} \\ \m{0} \end{bmatrix}. \label{equ:syst040}
\end{equation}
Hence, every column vector of $\breve{\m{G}}$ has to be orthogonal to the $N_u$ lowermost row vectors of $\m{F}_N^{-1}\m{B}\in\mathbb{C}^{N\times (N_d+N_r)}$. Note that $\m{F}_N^{-1}\m{B}$ is composed of those columns of $\m{F}_N^{-1}$ that correspond to the non-zero entries of the OFDM frequency domain symbol $\vef{x}$. \smallskip

Consequently, we have to solve a constrained optimization problem for $\breve{\m{G}}$, which can finally be written as 
\begin{equation}
  \breve{\m{G}}=\mathrm{argmin}\left\{J_\text{BLUE}\right\} \hspace{0.3cm} \mathrm{s.t.} \hspace{0.15cm} 																					\m{F}_N^{-1}\m{B}\breve{\m{G}} = \begin{bmatrix} \m{*} \\ \m{0} \end{bmatrix}. \label{equ:syst017}
\end{equation}
The solutions to the optimization problem will lead to code generator matrices $\breve{\m{G}}$ matched to the BLUE 'decoding' procedure. 

\subsection{Transceiver Cost Function for the LMMSE Data Estimator}
In this subsection we assume that the LMMSE estimator will be used for data estimation. Again we are aiming for optimizing the overall system performance by minimizing the sum over the error variances after the data estimation. Using \eqref{equ:LMMSE_006} with $\mf{H}=\m{I}$ this leads to 
\begin{equation}
	J = N\sigma_n^2 \mathrm{tr}\left\{(\breve{\m{G}}^H\breve{\m{G}}+ \frac{N\sigma_n^2}{\sigma_d^2}\m{I})^{-1}\right\}.  								\label{equ:syst00}
\end{equation}
Like in the considerations for the BLUE we fix $c=E_s/\sigma_n^2$, and we therefore insert \eqref{equ:syst004} into \eqref{equ:syst00} to obtain the cost function
\begin{equation}
	J_\text{LMMSE} = \frac{\sigma_d^2}{c N_d} \mathrm{tr}\{\breve{\m{G}}^H\breve{\m{G}}\}
											\mathrm{tr}\left\{\left(\breve{\m{G}}^H\breve{\m{G}}+
											\frac{\mathrm{tr}\{\breve{\m{G}}^H\breve{\m{G}}\}}{c N_d}\m{I}\right)^{-1}\right\}.  \label{equ:syst018}
\end{equation}
Alternatively, the cost function can also be written as
\begin{equation}
	J_\text{LMMSE} = \sigma_d^2 \mathrm{tr}\left\{\left(
										\frac{c N_d}{\mathrm{tr}\{\breve{\m{G}}^H\breve{\m{G}}\}} \breve{\m{G}}^H\breve{\m{G}}
											+\m{I}\right)^{-1}\right\}.  \label{equ:syst035}
\end{equation}
The constrained optimization problem to find $\breve{\m{G}}$ can finally be written as 
\begin{equation}
  \breve{\m{G}}=\mathrm{argmin}\left\{J_\text{LMMSE}\right\} \hspace{0.3cm} \mathrm{s.t.} \hspace{0.15cm} 																			\m{F}_N^{-1}\m{B}\breve{\m{G}} = \begin{bmatrix} \m{*} \\ \m{0} \end{bmatrix}. \label{equ:syst026}
\end{equation}
For sufficiently large $c$ we have $J_\text{LMMSE}\approx J_\text{BLUE}$, and the particular choice of $c$ is again irrelevant for the searching of optimum code generator matrices. However, this is not immediately apparent for small values of $c$. The solutions to the optimization problem will lead to code generator matrices $\breve{\m{G}}$ matched to the LMMSE 'decoding' procedure.

\section{Solutions of the Optimization Problems}\label{sec:optimization}
In this section we at first solve the optimization problems in \eqref{equ:syst017} and \eqref{equ:syst026} numerically. The solutions are ambiguous, however all found code generator matrices share a number of common properties which we will discuss in detail. With the help of analytical considerations which are partly shifted to the appendices we show that all found code generator matrices not only correspond to local minima but to the global minimum of the associated cost function.     

\subsection{Preparatory Steps and Numerical Solution with the Steepest Descent Algorithm}\label{sec:num}

In this section we use the steepest descent algorithm to numerically solve the optimization problems in \eqref{equ:syst017} and \eqref{equ:syst026}. As a preparatory step we transform the constrained optimization problems into unconstrained problems. For that and according to \eqref{equ:trans023} we write $\breve{\m{G}}$ in the form
\begin{equation}
	\breve{\m{G}} = \m{A}\m{P}\begin{bmatrix} \m{I} \\ \breve{\m{T}} \end{bmatrix}, \label{equ:syst002}
\end{equation}
with a non-singular real matrix $\m{A}\in\mathbb{R}^{(N_d+N_r) \times (N_d+N_r)}$, and with a fixed permutation matrix $\m{P}$ as e.g., found by minimizing the cost function in \eqref{equ:uw001}. The constraint in \eqref{equ:syst017} and \eqref{equ:syst026} can thus be rewritten as
\begin{equation}
	\m{F}_N^{-1}\m{B}\m{A}\m{P}\begin{bmatrix} \m{I} \\ \breve{\m{T}} \end{bmatrix} 
					= \begin{bmatrix} \m{*} \\ \m{0} \end{bmatrix}. \label{equ:sys0017}
\end{equation}
With the introduction of
\begin{equation}
	\breve{\m{M}}=\m{F}_N^{-1} \m{B} \m{A}\m{P}= \begin{bmatrix} \breve{\m{M}}_{11} & \breve{\m{M}}_{12} \\ 
																								\breve{\m{M}}_{21} & \breve{\m{M}}_{22}\end{bmatrix}, \label{equ:syst010}
\end{equation}
the constraint in \eqref{equ:sys0017} can now simply be fulfilled by choosing $\breve{\m{T}}$ as 
\begin{equation}
	\breve{\m{T}} = -(\breve{\m{M}}_{22})^{-1}\breve{\m{M}}_{21}. \label{equ:syst011}
\end{equation}
That means that for a given non-singular real matrix $\m{A}$, the matrices $\breve{\m{T}}$ and $\breve{\m{G}}$ can un-ambiguously be calculated by \eqref{equ:syst011} and \eqref{equ:syst002}, respectively, such that the constraint in \eqref{equ:syst017} and \eqref{equ:syst026} is automatically fulfilled. We can therefore consider the cost functions $J_\text{BLUE}$ and $J_\text{LMMSE}$ as functions of the real matrix $\m{A}$, where $\breve{\m{T}}$ and $\breve{\m{G}}$ have to be determined by \eqref{equ:syst011} and \eqref{equ:syst002}, respectively. The steepest descent algorithm can then be applied to the unconstrained optimization problems
\begin{equation}
	\m{A}_\text{opt}=\mathrm{argmin}\left\{J_\text{BLUE, LMMSE}\right\} \label{equ:syst021}
\end{equation}
in a straight forward manner. In \eqref{equ:syst021} either $J_\text{BLUE}$ or $J_\text{LMMSE}$ is minimized. By following this approach the steepest descent algorithm automatically only searches within a subset of matrices $\breve{\m{G}}$ that fulfill the constraint in \eqref{equ:syst017} and \eqref{equ:syst026}, or in other words that produce a zero UW in the OFDM time domain symbols. \smallskip

For the steepest descent algorithm the gradients of the cost functions $J_\text{BLUE}$ and $J_\text{LMMSE}$ with respect to the real matrix $\m{A}$ are required. We approximated the partial derivations $\partial J / \partial [A]_{ij}$ by
\begin{equation}
	\frac{\partial J}{\partial [A]_{ij}} = \frac{J([A]_{ij}+\epsilon)-J([A]_{ij}-\epsilon)}{2\epsilon},
\end{equation}
with a very small $\epsilon$. For $J$ we inserted $J_\text{BLUE}$ or $J_\text{LMMSE}$, respectively. We used two different approaches for the initialization of the steepest descent algorithm: \smallskip

\subsubsection{Initialization with the Code Generator Matrix $\m{G}$} \label{seq:init1}
In our first approach we chose the initialization 
\begin{equation}
	\m{A}^{(0)}=\m{I} \label{equ:init1}
\end{equation}
which implies $\breve{\m{T}}^{(0)}=\m{T}$ and 
\begin{equation}
	\breve{\m{G}}^{(0)}=\m{P}\begin{bmatrix} \m{I} & \m{T}^T \end{bmatrix}^T = \m{G}. 
\end{equation}
The iterative optimization process consequently starts with the code generator matrix $\m{G}$ of our original systematic coded UW-OFDM concept, which can definitely be assumed to be a good initial guess. We denote the resulting optimum code generator matrix (found after convergence of the algorithm) with $\breve{\m{G}}'$.

\subsubsection{Random Initialization} \label{seq:init2}
In the second approach we chose each element of $\m{A}^{(0)}$ as a realization of a Gaussian random variable with mean zero and variance one:
\begin{equation}
	[\m{A}^{(0)}]_{ij} \sim \mathcal{N}(0,1) \label{equ:init2}
\end{equation}
We denote the resulting code generator matrix with $\breve{\m{G}}''$.\bigskip

For both cost functions $J_\text{BLUE}$ and $J_\text{LMMSE}$ we can claim the
following: By using the initialization as in \eqref{equ:init1} the steepest
descent algorithm converges at least one order of magnitude faster compared to
the case when \eqref{equ:init2} is used. For the random initialization approach
the resulting code generator matrix generally varies from trial to trial.

\subsection{General Properties of Optimum Code Generator Matrices for the BLUE} \label{sec:BLUE_optimization}
Interestingly, all found local minima for the BLUE based numerical optimization feature the same value of the cost function $J_\text{BLUE,min}$, independently of the choice of the initialization $\m{A}^{(0)}$. Another highly interesting finding is, that all resulting code generator matrices (again independently of $\m{A}^{(0)}$) feature the property
\begin{equation}
	\breve{\m{G}}^H\breve{\m{G}}=\alpha \m{I} \label{equ:syst024}
\end{equation}
with some constant $\alpha$ (which may vary dependent on the results of the optimization process). This property has a number of important implications. First, inserting \eqref{equ:syst024} into the cost function \eqref{equ:syst007} leads to
\begin{equation}
  J_\text{BLUE,min} = \frac{\sigma_d^2}{c N_d} (N_d\alpha) (N_d \alpha^{-1}) = \frac{\sigma_d^2 N_d}{c}, \label{equ:syst030}
\end{equation}
which is in agreement with the numerically found local minima. We can conclude that every $\breve{\m{G}}$ fulfilling
\begin{equation}
	\breve{\m{G}}^H\breve{\m{G}}=\alpha \m{I} \hspace{0.3cm} \mathrm{and} \hspace{0.15cm} \m{F}_N^{-1}\m{B}\breve{\m{G}} =																						\begin{bmatrix} \m{*} \\ \m{0} \end{bmatrix}
																\label{equ:syst025}
\end{equation}
for any value of $\alpha$ will also result in the same value $J_\text{BLUE,min}$ of the cost function, and will produce a zero UW in time domain. Second, if we apply a code generator matrix satisfying \eqref{equ:syst025}, then the error covariance matrix after the data estimation in the AWGN channel is given by 
\begin{equation}
	\m{C}_{\tilde{e}\tilde{e},\text{BLUE}} = \frac{\sigma_d^2}{c}\m{I}.   \label{equ:syst070}    			              
\end{equation}
This simply follows from inserting \eqref{equ:syst024} and \eqref{equ:syst004} into \eqref{equ:ZF_009}. As an important consequence we can conclude that the noise at the output of the BLUE is uncorrelated under AWGN conditions. This is clearly in contrast to systematic coded UW-OFDM, where $\m{C}_{\tilde{e}\tilde{e}}$ is non-diagonal also in the AWGN channel case. And third, \eqref{equ:syst024} implies that all singular values of $\breve{\m{G}}$ are identical. To show this we consider a singular value decomposition (SVD) of $\breve{\m{G}}$ as
\begin{equation}
	\breve{\m{G}} = \m{U} \m{\Sigma} \m{V}^H, \label{equ:an001}
\end{equation}
with unitary matrices $\m{U}$ and $\m{V}$, and with the matrix $\m{\Sigma}=\begin{bmatrix} \m{D} & \m{0} \end{bmatrix}^T$, where $\m{D}$ is a real diagonal matrix having the singular values $s_1,s_2,...,s_{N_d}$ of $\breve{\m{G}}$ at its main diagonal. With \eqref{equ:syst024} we therefore have 
\begin{align}
	\alpha\m{I} &= \breve{\m{G}}^H\breve{\m{G}} = \m{V}\m{\Sigma}^H\m{U}^H\m{U}\m{\Sigma}\m{V}^H = \m{V}\m{D}^2\m{V}^H  \nonumber \\
	\Leftrightarrow \alpha\m{I}	&= \m{D}^2=\text{diag}\left\{s_1^2,s_2^2,...,s^2_{N_d}\right\}.  \label{equ:num010}
\end{align}
From \eqref{equ:num010} it follows that $\breve{\m{G}}^H\breve{\m{G}}=\alpha\m{I}$ implies $\alpha=s_1^2=s_2^2=\cdots s^2_{N_d}:=s^2$. The property in \eqref{equ:syst024} can therefore also be written as 
\begin{equation}
	\breve{\m{G}}^H\breve{\m{G}}=s^2 \m{I}. \label{equ:syst045}
\end{equation}
The argumentation can also be done the other way round: If all singular values of $\breve{\m{G}}$ are identical then we have $\breve{\m{G}}^H\breve{\m{G}}=\alpha\m{I}$ with $\alpha = s^2$. 

An open question is still whether the value of the cost function
$J_\text{BLUE,min}$ as in \eqref{equ:syst030} corresponding to the numerically
found local minima depicts the global minimum of the constrained optimization problem in
\eqref{equ:syst017}. To answer this question we now at first merely concentrate
on the cost function $J_\text{BLUE}$, and we disregard the constraint in
\eqref{equ:syst017} for a moment: Let $\ve{s}=\begin{bmatrix} s_1 & s_2 &
\cdots & s_{N_d} \end{bmatrix}^T$ be the vector of singular values of
$\breve{\m{G}}$. In Appendix \ref{sec:appendix_a} we will analytically show
that $\partial J_\text{BLUE}/ \partial \ve{s}=\ve{0}$ if and only if all
singular values of $\breve{\m{G}}$ are identical. Consequently, every possible
candidate $\breve{\m{G}}$ for a local minimum satisfies
$\breve{\m{G}}^H\breve{\m{G}} = s^2\m{I}$ (cf. \eqref{equ:num010} and its
implications). Inserting $\breve{\m{G}}^H\breve{\m{G}} = s^2\m{I}$ into the
cost function \eqref{equ:syst007} leads to the same expression as in
\eqref{equ:syst030}, and hence, every $\breve{\m{G}}$ fulfilling
$\breve{\m{G}}^H\breve{\m{G}} = s^2\m{I}$ results in the same (and minimum)
value $J_\text{BLUE,min} = \sigma_d^2 N_d/c$ which therefore constitutes the
global minimum of the cost function.

We now come back to the constrained problem in \eqref{equ:syst017}: From our numerical solutions we know that matrices exist, that firstly satisfy $\breve{\m{G}}^H\breve{\m{G}} = s^2\m{I}$ and therefore result in the global minimum of the cost function $J_\text{BLUE}$, and that secondly fulfill the constraint $\m{F}_N^{-1}\m{B}\breve{\m{G}} =	\begin{bmatrix} \m{*} \\ \m{0} \end{bmatrix}$. With these considerations we finally end up with the following important proposition: \newline

\noindent \textbf{Properties of optimum code generator matrices: }\textit{A code generator matrix $\breve{\m{G}}$ is optimum, i.e., leads to a global minimum of the constrained optimization problem in} \eqref{equ:syst017}, \textit{if and only if $\breve{\m{G}}$ satisfies} 
\begin{align}
		\breve{\m{G}}^H\breve{\m{G}}		&= s^2 \m{I}  \hspace{0.3cm} \textit{and} \label{equ:syst060} \\
	 	\m{F}_N^{-1}\m{B}\breve{\m{G}} 	&=	\begin{bmatrix} \m{*} \\ \m{0} \end{bmatrix}, \label{equ:syst061}
\end{align}
\textit{where} $s:= s_1 = s_2 = \cdots =s_{N_d}$ \textit{are the (all identical) singular values of} $\breve{\m{G}}$. \textit{The global minimum of the cost function is given by \eqref{equ:syst030}, and the error covariance matrix after data estimation (in the AWGN channel) is the scaled identity matrix as given in \eqref{equ:syst070}.}  

\bigskip

Note that because of \eqref{equ:syst060} the colums of any optimum code generator matrix $\breve{\m{G}}$ form an orthogonal basis of an $N_d$-dimensional subspace of $\mathbb{C}^{(N_d+N_r)\times 1}$. Furthermore, as already discussed above, \eqref{equ:syst061} implies that every column vector of an optimum $\breve{\m{G}}$ is orthogonal to the $N_u$ lowermost row vectors of $\m{F}_N^{-1}\m{B}$.

\subsection{General Properties of Optimium Code Generator Matrices for the LMMSE Estimator}
In Appendix \ref{sec:appendix_b} we will analytically show that $\partial J_\text{LMMSE}/ \partial \ve{s}=\ve{0}$ if and only if all singular values of $\breve{\m{G}}$ are identical. All other findings from Sec. \ref{sec:BLUE_optimization} also hold for the LMMSE estimator based transceiver optimization, except the particular expressions for $J_\text{LMMSE,min}$ and $\m{C}_{\tilde{e}\tilde{e},\text{LMMSE}}$ differ slightly. With \eqref{equ:syst024}, \eqref{equ:LMMSE_006} and \eqref{equ:syst035} it immediately follows that
\begin{eqnarray}
  J_\text{LMMSE,min} &=& \frac{\sigma_d^2 N_d}{c+1},  \label{equ:syst014} \\
	\m{C}_{\tilde{e}\tilde{e},\text{LMMSE}} &=& \frac{\sigma_d^2}{c+1} \m{I}.   \label{equ:syst071}  			             
\end{eqnarray}
As a consequence of the above findings we learn that a code generator matrix which is optimum
for the BLUE based 'decoding' procedure is automatically also optimum for the
LMMSE based data estimation (and vice versa).

\subsection{Normalized Optimum Code Generator Matrices}
From \eqref{equ:syst030} and \eqref{equ:syst014} we learn that the particular value of $\alpha = s^2$ does not play any important role. In our simulations, cf. Sec. \ref{sec:simulations}, we therefore normalized all found code generator matrices such that $\alpha = s^2 = 1$ or $\breve{\m{G}}^H\breve{\m{G}}=\m{I}$. The columns of any normalized optimum code generator matrix $\breve{\m{G}}$ form an orthonormal basis of an $N_d$-dimensional subspace of $\mathbb{C}^{(N_d+N_r)\times 1}$. As another consequence of $s = 1$ the operation $\vef{c}=\breve{\m{G}}\vef{d}$ becomes energy-invariant and we have $E_\vef{c}=E_\vef{d}=N_d\sigma_d^2$, and the mean energy of an OFDM time domain symbol (for the zero UW case) follows to $E_\ve{x} = N_d\sigma_d^2/N$; cf. \eqref{equ:syst041}. \smallskip

\subsection{Comparison of Generator Matrices obtained from different Initialization Strategies} \label{seq:matrices}
We will now discuss some further interesting properties of two particular numerically found solutions. Here we only concentrate on $\breve{\m{G}}'$ and on one particular solution for $\breve{\m{G}}''$ (and the corresponding matrices $\m{A}_\text{opt}'$ and $\m{A}_\text{opt}''$) found by applying the LMMSE estimator based optimization with $c=1$ for the initializations as described in \eqref{equ:init1} and \eqref{equ:init2}, respectively.  \smallskip

The matrix $\m{A}_\text{opt}'$ features the symmetry property 
\begin{equation}
	\m{A}_\text{opt}' = [ \ve{a}_0' \cdots \ve{a}_{N_a/2-1}' \hspace{0.2cm} \mathrm{flip}\{\ve{a}_{N_a/2-1}'\} \cdots \mathrm{flip}\{\ve{a}_0'\}], \label{equ:syst008}
\end{equation}
where $N_a = N_d+N_r$ and the $\ve{a}_i'$ with $i=0,1,...,N_a/2-1$ are the first $N_a/2$ colums of $\m{A}_\text{opt}'$. The corresponding code generator matrix $\breve{\m{G}}'$ shows the same symmetry property as $\m{G}$ in Section \ref{sec:uw}, namely
\begin{equation}
	\breve{\m{G}}' = [ \breve{\ve{g}}_0' \cdots  \breve{\ve{g}}_{N_d/2-1}' \hspace{0.06cm} 
										\mathrm{flip}\{(\breve{\ve{g}}_{N_d/2-1}')^{*}\} \cdots 	\mathrm{flip}\{(\breve{\ve{g}}_0')^{*}\}]. 																	\label{equ:syst029}
\end{equation}
Here the $\breve{\ve{g}}_i'$ with $i=0,1,...,N_d/2-1$ are the first $N_d/2$ columns of $\breve{\m{G}}'$. These symmetry properties do not hold for $\m{A}_\text{opt}''$ and $\breve{\m{G}}''$, respectively. It appears that the matrices $\m{A}_\text{opt}'$ and $\m{A}_\text{opt}''$ show a completely different construction. $\m{A}_\text{opt}'$ approximately features a band matrix structure, where all dominant entries are positioned on the main diagonal and the first few diagonals directly above and below the main diagonal. The remaining elements are close to zero. In contrast $\m{A}_\text{opt}''$ is a full matrix. This results in completely different (pre)coding properties of $\breve{\m{G}}'$ and $\breve{\m{G}}''$. $\breve{\m{G}}'$ can be regarded as the natural perfecting of $\m{G}$ and is constructed such that the energy of one data symbol is mainly (however, not exclusively) spread locally. In contrast $\breve{\m{G}}''$ spreads the energy of each data symbol approximately uniformly over the codeword $\vef{c}$. While $\breve{\m{G}}'$ and $\breve{\m{G}}''$ perform identically in an AWGN environment the discussed differences lead to a quite different behavior of $\breve{\m{G}}'$ and $\breve{\m{G}}''$ in frequency selective environments. We will exemplify these differences and the consequences in the simulation section. \smallskip

\begin{figure}[!ht]
\centering
\includegraphics[width=3.5in]{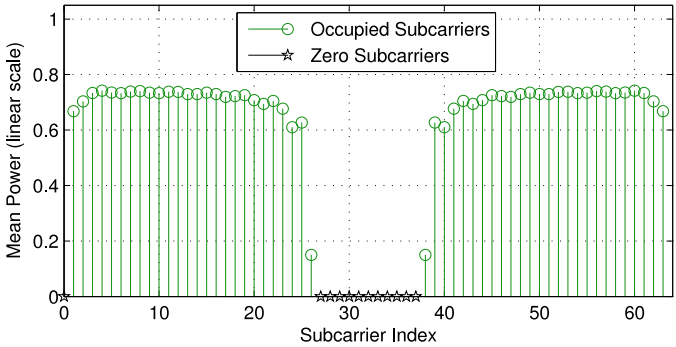}
\caption{Mean power of individual subcarrier symbols for Example 2.}
\label{fig:mean_power3}
\end{figure}

\noindent 
\textit{Example 2:} We apply the system parameters as in Example 1 and investigate the mean power levels for all individual subcarriers for the case the non-systematic code generator matrix $\breve{\m{G}}'$ is applied. The mean power values for the codeword symbols correspond to the diagonal elements of the covariance matrix $\m{C}_{\tilde{c}\tilde{c}}=\sigma_d^2\breve{\m{G}}'(\breve{\m{G}}')^H$. Fig. \ref{fig:mean_power3} shows the power distribution over all subcarrier symbols (additionally also including the zero subcarrier symbols) again for the case the UW is the zero word $\ve{x}_u=\ve{0}$ and for $\sigma_d^2=1$.  
We can clearly identify that our chosen optimality criterion also implies a significant reduction of the power levels of the former redundant subcarriers compared to the original UW-OFDM approach; cf. Fig. \ref{fig:mean_power}. Furthermore, it can be seen that the redundant energy is now smeared over all subcarriers; consequently we can no longer speak of data or redundant subcarriers. The power levels are quite similar for all symbols, the only exceptions are the two subcarrier symbols at the band edges. Due to the normalization of $\breve{\m{G}}'$ such that $(\breve{\m{G}}')^H\breve{\m{G}}'=\m{I}$ we have  $\mathrm{tr}\{\m{C}_{\tilde{c}\tilde{c}}\} = \sigma_d^2 \mathrm{tr}\{\breve{\m{G}}'(\breve{\m{G}}')^H\} = \sigma_d^2 \mathrm{tr}\{(\breve{\m{G}}')^H\breve{\m{G}}'\}=N_d\sigma_d^2$. In Fig. \ref{fig:mean_power3} the sum over all mean power levels is therefore 36.

\section{On the Relationship between UW-OFDM and UW-SC/FDE} \label{sec:scfde}
With the help of a specific constructed code generator matrix non-systematic coded UW-OFDM can be converted into a UW-SC/FDE system. For that we assume for the moment that no zero subcarriers are used, or $\m{B}=\m{I} \in \mathbb{R}^{N\times N}$ and $N_d=N-N_r$, and we consider the matrix
\begin{equation}
	\breve{\m{G}}_\text{SC} = \m{F}_N\begin{bmatrix} \m{I} \\ \m{0} \end{bmatrix}.
\end{equation}
$\breve{\m{G}}_\text{SC}$ fulfills
\begin{equation}
	\breve{\m{G}}_\text{SC}^H\breve{\m{G}}_\text{SC}=N \m{I} \hspace{0.3cm} \mathrm{and} \hspace{0.15cm} 																			\m{F}_N^{-1}\m{B}\breve{\m{G}}_\text{SC} = \begin{bmatrix} \m{I} \\ \m{0} \end{bmatrix},
						\label{equ:sc001}
\end{equation}
and consequently constitutes an optimum code generator matrix; cf. \eqref{equ:syst060} and \eqref{equ:syst061}. Furthermore, it is apparent that $\breve{\m{G}}_\text{SC}$ generates a UW-SC/FDE signal with a zero UW $\ve{x}_u=\ve{0}$ since the time domain symbol vector follows to
\begin{equation}
	\ve{x} = \m{F}_N^{-1}\m{B}\breve{\m{G}}\vef{d} = \begin{bmatrix} \m{I} \\ \m{0} \end{bmatrix} \vef{d} = \begin{bmatrix} \vef{d} \\ \ve{0} \end{bmatrix}.
\end{equation} 
Simulation results for a UW-SCFDE system in comparison with systematic coded UW-OFDM can be found in \cite{Huemer10_2}.    

\section{Simulation Results} \label{sec:simulations}
In this section we present a number of simulation results to show the
advantageous features of the developed non-systematic coded UW-OFDM concept. In
our simulations the transmitter processing starts with optional (outer) channel
coding, interleaving and QAM-mapping (we apply QPSK symbols unless specified
otherwise). We used the same outer convolutional encoder as defined in
\cite{IEEE99}, and we show results for (outer) coding rates $r=3/4$ and
$r=1/2$, respectively. Next the complex codewords are determined by either
using $\m{G}$, $\breve{\m{G}}'$ or $\breve{\m{G}}''$. Here, $\breve{\m{G}}'$
and $\breve{\m{G}}''$ are the particular code generator matrices dicussed in
Sec. \ref{seq:matrices}. We note that the (optional) outer convolutional code
is a binary code while the inner code described by $\m{G}$, $\breve{\m{G}}'$ or
$\breve{\m{G}}''$ is an RS code over the field of complex numbers. The latter
is naturally always inherently present due to the proposed way of generating
UW-OFDM symbols with zero UWs at their tails. After applying a code generator
matrix, zero subcarriers are filled in, and the IFFT (inverse fast Fourier
transform) is performed. Finally, the desired UW is added in time domain. At
the receiver side the processing for one OFDM symbol starts with an FFT, then
the influence of the UW ($\mf{H}\m{B}^T \vef{x}_u$) is subtracted;
cf. \eqref{equ:system_model_002}-\eqref{equ:system_model_001}. Next the data
estimation is applied. Finally demapping, deinterleaving and (outer) channel
decoding are performed. For the applied soft decision Viterbi channel decoder
the main diagonal of the appropriate matrix $\m{C}_{\tilde{e}\tilde{e}}$ is
used to specify the (in case of transmitting over frequency selective channels)
varying noise variances along the subcarriers after data estimation. \smallskip

\subsection{Simulation Setup}
We compare our UW-OFDM approaches with the classical CP-OFDM concept. The IEEE 802.11a WLAN standard \cite{IEEE99} serves as reference system. We apply the same parameters for UW-OFDM as in \cite{IEEE99} wherever possible, the most important parameters used in our simulations are specified in Table \ref{tab2}. 
\renewcommand{\arraystretch}{1.2}
\begin{table} [htb]
\caption{\label{tab2} Main PHY parameters of the investigated systems.}
\begin{center}
\begin{tabular}  {|l|c|c|} \hline
														  	& 802.11a				& UW-OFDM       \\ \hline\hline
Modulation schemes            	& QPSK, 16QAM
& QPSK, 16QAM 					\\ \hline
Coding rates (outer code)     	& 1/2, 3/4 			& 1/2, 3/4			\\ \hline
Occupied subcarriers					  & 52  					& 52						\\ \hline
Data subcarriers     						& 48 						&	36						\\ \hline
Additional subcarriers 			  	& 4 (pilots) 		&  16 (redundant) \\ \hline
DFT period								   		& 3.2 $\mu$s 		&  3.2 $\mu$s		\\ \hline
Guard duration 									& 800 ns (CP)		&  800 ns (UW)	\\ \hline
Total OFDM symbol duration    	& 4 $\mu$s 			& 3.2 $\mu$s		\\ \hline
Subcarrier spacing							& 312.5 kHz     & 312.5 kHz			\\ \hline
\end{tabular}
\end{center}
\end{table}
The sampling frequency has been chosen to be $f_s = 20$MHz. As in \cite{IEEE99} the indices of the zero subcarriers within an OFDM symbol $\vef{x}$ are set to \{0, 27, 28,...,37\}. The indices of the redundant subcarriers are chosen to be \{2, 6, 10, 14, 17, 21, 24, 26, 38, 40, 43, 47, 50, 54, 58, 62\} as already discussed in the Example 1 in Sec. \ref{sec:uw}. Note that in conventional CP-OFDM like in the WLAN standard, the total length of an OFDM symbol is given by $T_{DFT}+T_{GI}$. However, the guard interval is part of the DFT period in the UW-OFDM approach which leads to significantly shorter total symbol durations. It is therefore important to mention that the compared systems show (almost) identical bandwidth efficiencies. \smallskip

Note that in the IEEE 802.11a standard 4 pilot subcarriers are specified. Those are used for estimation and synchronization purposes at the receiver side. In our UW-OFDM approaches we omitted these pilots, because the unique word, which is deterministic, shall (at least) take over the estimation and synchronization tasks which are normally performed with the help of the 4 pilot subcarriers. In order to make a fair BER performance comparison, the energy of the UW related to the total mean energy of a transmit symbol is set to 4/52 in our BER simulations. This exactly corresponds to the total energy of the 4 pilots
related to the total mean energy of a transmit symbol in the IEEE standard. As UW we applied a linear chirp sequence exhibiting the same bandwidth as the data signal, and featuring a constant envelope in time domain and approximately a constant envelope in frequency domain. However, the particular shape of the UW has no impact on the BER behavior; cf. \cite{Onic10_1}.

\begin{figure}[t]
\centering
\includegraphics[width=3.5in]{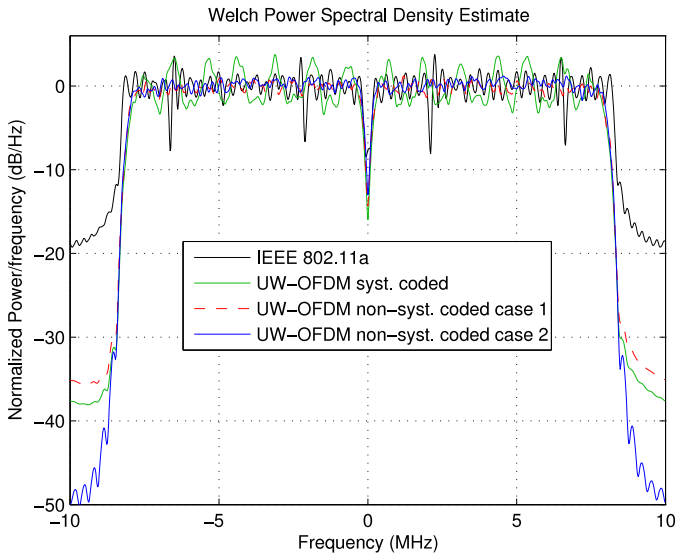}
\caption{Power spectral density comparison: CP-OFDM, UW-OFDM using $\m{G}$ (systematic coded), $\breve{\m{G}}'$ (non-systematic coded - case 1) and $\breve{\m{G}}''$ (non-systematic coded - case 2).}
\label{fig:PSD}
\end{figure}

\subsection{Power Spectral Density}
Fig. \ref{fig:PSD} shows the estimated power spectral densities (PSDs) of
simulated UW-OFDM bursts and of a CP-OFDM burst. For all cases we simulated a
burst composed of a preamble (in all cases the IEEE 802.11a preamble), and a
data part consisting of 1000 bytes of data. We used an outer channel code with
coding rate $r=1/2$. For the UW-OFDM concepts ($\m{G}$, $\breve{\m{G}}'$,
$\breve{\m{G}}''$) we exceptionally applied the zero UW for these PSD investigations. Note that we did not use any additional filters for spectral shaping. Fig. \ref{fig:PSD} clearly shows that the UW-OFDM spectra feature a significantly better sidelobe suppression compared to the CP-OFDM spectrum. The out-of-band emissions generated by $\m{G}$ and $\breve{\m{G}}'$ are more than 15dB below the emissions of the CP-OFDM system. The emissions are even notably lower for $\breve{\m{G}}''$. Furthermore, the spectra for $\breve{\m{G}}'$ and $\breve{\m{G}}''$ feature an extremely flat in-band region compared to systematic coded UW-OFDM. This can be explained by the fact, that for the systematic coded case the mean power strongly varies between data and redundant subcarriers, cf. Fig. \ref{fig:mean_power}, while all subcarriers (except the ones at the band edges) show almost equal power in the non-systematic case.

\subsection{BER Simulation Results with Perfect Channel Knowledge}
We will now show BER simulation results for the AWGN channel as well as for frequency selective environments. To avoid confusions in the figures we at first only use $\breve{\m{G}}'$ for the non-systematic coded UW-OFDM system, and at the end of the section we will then also show and interpret results for $\breve{\m{G}}''$. Perfect channel knowledge at the receiver is assumed in all simulations. \smallskip

\subsubsection{Results for the AWGN case}
\begin{figure}[t]
\centering
\includegraphics[width=3.5in]{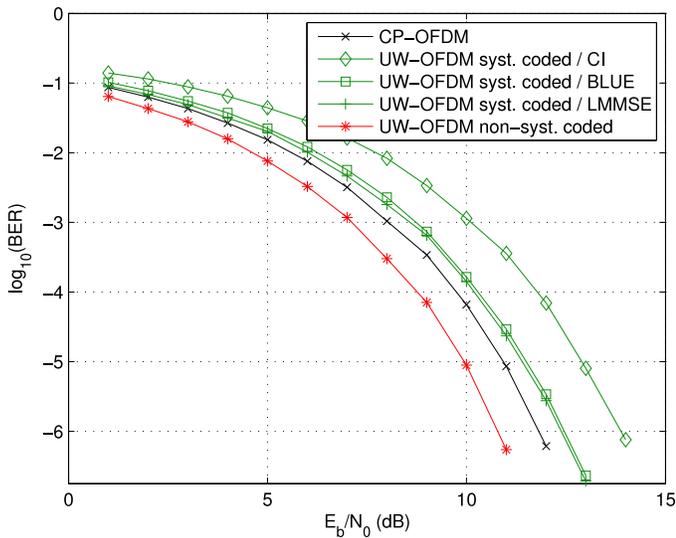}
\caption{BER comparison between the UW-OFDM approaches ($\m{G}$ and
  $\breve{\m{G}}'$) and the CP-OFDM system in the AWGN channel (QPSK).}
\label{fig:BER_QPSK_AWGN}
\end{figure}

Clearly, OFDM is designed for data transmission in frequency selective environments. Nevertheless, we start our comparison with simulation results in the AWGN channel, since we optimized the non-systematic code generator matrices for that case. In Fig. \ref{fig:BER_QPSK_AWGN} the BER behavior of the IEEE 802.11a CP-OFDM based standard, and of both, the systematic coded ($\m{G}$) and the non-systematic coded ($\breve{\m{G}}'$) UW-OFDM approach are compared under AWGN conditions. No outer code is used for these simulations. 
Simulation results are provided for the BLUE and for the LMMSE data estimator,
respectively. For the systematic coded system additional results for a simple
channel inversion (CI) receiver ($\m{E}=\begin{bmatrix} \m{I} & \m{0}
\end{bmatrix} \mf{H}^{-1}$) are included for comparison reasons. For the
systematic coded UW-OFDM system it can be observed that the BLUE and the LMMSE
estimator clearly outperform the CI receiver, and the LMMSE estimator performs
slightly better than the BLUE. For non-systematic coded UW-OFDM we only plotted
one single curve since the BLUE and the LMMSE data estimator perform completely
identical. This is somewhat surprising since the error variances after data estimation are not identical; cf. \eqref{equ:syst070} and \eqref{equ:syst071}. However, we found that in the AWGN case the QPSK symbol estimates of the BLUE and of the LMMSE data estimator always lie in the same decision region of the constellation diagram, and the difference in the error variances does not translate into a difference in the BER performance. To give some numbers we compare the performances at a bit error ratio of $10^{-6}$. Systematic coded UW-OFDM performs slightly worse compared to the CP-OFDM reference system, the non-systematic coded UW-OFDM system outperforms CP-OFDM by 1dB and systematic coded UW-OFDM (with LMMSE data estimation) by 1.6dB, respectively. We consider this as a remarkable performance of the novel non-systematic coded UW-OFDM system. \bigskip

\subsubsection{Results for frequency selective environments ($\m{G}$, $\breve{\m{G}}'$)}
We now turn to results in frequency selective indoor environments. Since the LMMSE estimator always outperforms the BLUE in dispersive channels we only concentrate on the LMMSE estimator in the following. For the simulation of indoor multipath channels we applied the model described in \cite{Fak97}, which has also been used during the IEEE 802.11a standardization process. The channel impulse responses are modeled as tapped delay lines, each tap with uniformly distributed phase and Rayleigh distributed magnitude, and with power decaying exponentially. The model allows the choice of the channel delay spread. For the following simulations we have generated and stored 5000 realizations of
channel impulse responses, all featuring a delay spread of 100ns and a total
length not exceeding the guard interval duration. Furthermore, the channel
impulse responses have been normalized such that the receive power is
independent of the actual channel. The subsequent figures represent BER results
averaged over that 5000 channel realizations. \smallskip

\begin{figure}[t]
\centering
\includegraphics[width=3.5in]{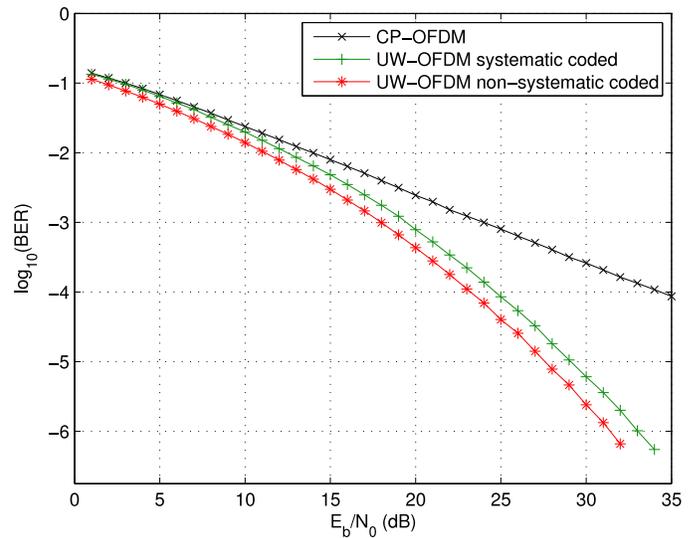}
\caption{BER comparison between the UW-OFDM approaches ($\m{G}$ and
  $\breve{\m{G}}'$) and the CP-OFDM system in a frequency selective environment
  w/o outer coding (QPSK).}
\label{fig:BER_QPSK_5000channels_uncoded}
\end{figure}

We start with simulation results for the case no outer code is used; cf. Fig. \ref{fig:BER_QPSK_5000channels_uncoded}. 
The gain achieved by systematic coded UW-OFDM over CP-OFDM is already
remarkable. Besides the coding gain achieved by $\m{G}$ together with the LMMSE
data estimator this mainly comes from the fact, that due to the dispersive preprocessing
property of $\m{G}$ data symbols corresponding to deep fading holes in the
channel's frequency response can still be detected reasonably, since portions
of these data symbols are also available at redundant subcarriers. Further, the
non-systematic coded UW-OFDM system outperforms the systematic coded one by
another 1.6dB, even though $\breve{\m{G}}'$ has been optimized for the AWGN
channel case. \smallskip

\begin{figure}[t]
\centering
\includegraphics[width=3.5in]{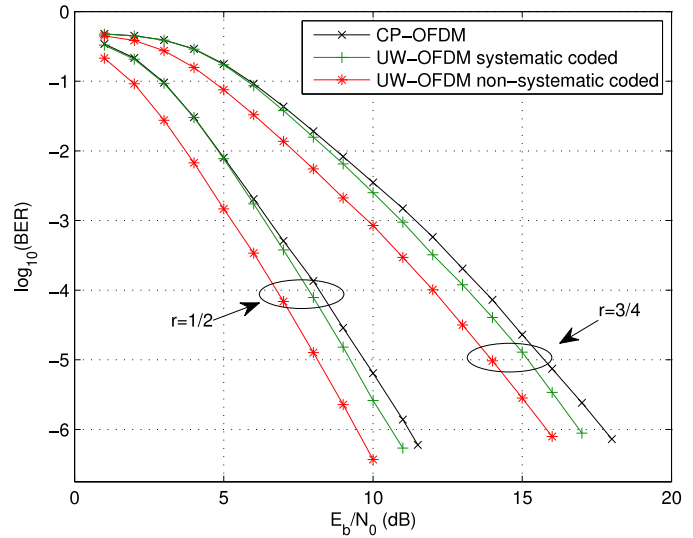}
\caption{\footnotesize{BER comparison between the UW-OFDM approaches ($\m{G}$
    and $\breve{\m{G}}'$) and the CP-OFDM system in a frequency selective
    environment with additional outer coding (QPSK).}}
\label{fig:BER_QPSK_5000channels}
\end{figure}

Next we present simulation results for the case the additional outer channel code is used; cf. Fig. \ref{fig:BER_QPSK_5000channels}. 
For both outer coding rates the UW-OFDM approaches outperform CP-OFDM, and
non-systematic coded UW-OFDM shows by far the best performance. The gains of
non-systematic coded UW-OFDM over CP-OFDM at a bit error ratio of $10^{-6}$ are
1.9dB and 1.7dB for $r=\tfrac{3}{4}$ and $r=\tfrac{1}{2}$, respectively, the
gains over systematic coded UW-OFDM are 1.1dB for both coding rates. \smallskip

\begin{figure}[t]
\centering
\includegraphics[width=3.5in]{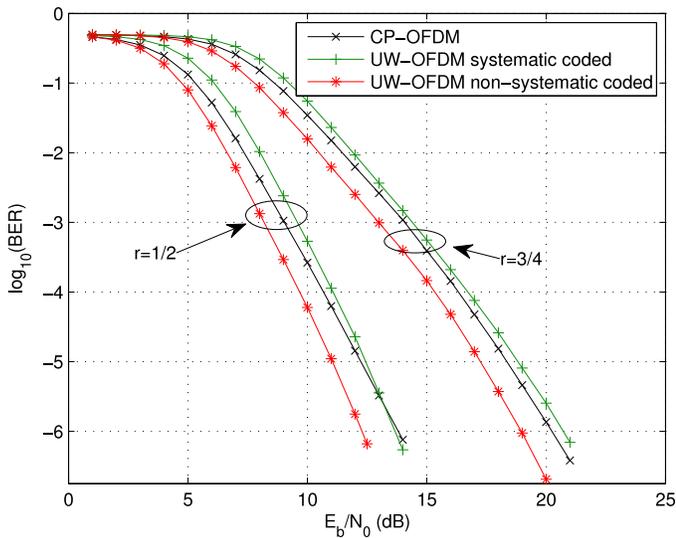}
\caption{\footnotesize{BER comparison between the UW-OFDM approaches ($\m{G}$
    and $\breve{\m{G}}'$) and the CP-OFDM system in a frequency selective
    environment with additional outer coding and 16QAM as modulation alphabet.}}
\label{fig:BER_16QAM_5000channels}
\end{figure}

Similar tendencies can also be observed in case 16QAM symbols are applied as
modulation alphabet; cf. Fig. \ref{fig:BER_16QAM_5000channels}. Non-systematic
coded UW-OFDM again significantly outperforms CP-OFDM by 1.6dB and 1.3dB for
$r=\tfrac{1}{2}$ and $r=\tfrac{3}{4}$, respectively (again measured at a bit
error ratio of $10^{-6}$). However, the gain of systematic coded UW-OFDM over
CP-OFDM shrinks to 0.2dB for $r=\tfrac{1}{2}$, and even turns to a loss of
0.5dB for $r=\tfrac{3}{4}$. Consequently and quite remarkably, the achieved
gain of $\breve{\m{G}}'$ over $\m{G}$ turns out to be notably larger in case of
16QAM compared to QPSK modulation.
\bigskip 

\subsubsection{Results for frequency selective environments ($\breve{\m{G}}'$, $\breve{\m{G}}''$)}
Finally we compare the performance of the two different derived code generator matrices $\breve{\m{G}}'$ and $\breve{\m{G}}''$. We can state that in the AWGN channel they feature exactly the same performance. This was expected since every optimum code generator matrix shows the same error covariance matrix for the AWGN case. However, they feature quite a different behavior in dispersive channels. In Sec. \ref{seq:matrices} we already discussed the different structures of $\breve{\m{G}}'$ and the particularly chosen $\breve{\m{G}}''$. We remind the reader that $\breve{\m{G}}'$ is constructed such that the energy of one data symbol is mainly (however, not exclusively) spread locally. $\breve{\m{G}}'$ can be regarded as the natural perfecting of $\m{G}$. In contrast $\breve{\m{G}}''$ spreads the energy of each data symbol approximately uniformly over the codeword $\vef{c}$. From this point of view the system with $\breve{\m{G}}''$ behaves comparable to a single carrier system, where the energy of each individual QAM symbol is also approximately uniformly distributed over the whole bandwidth. In contrast the system with $\breve{\m{G}}'$ still rather shows more similarity to classical OFDM, where a subcarrier exactly corresponds to one QAM symbol. 
\begin{figure}[t]
\centering
\includegraphics[width=3.5in]{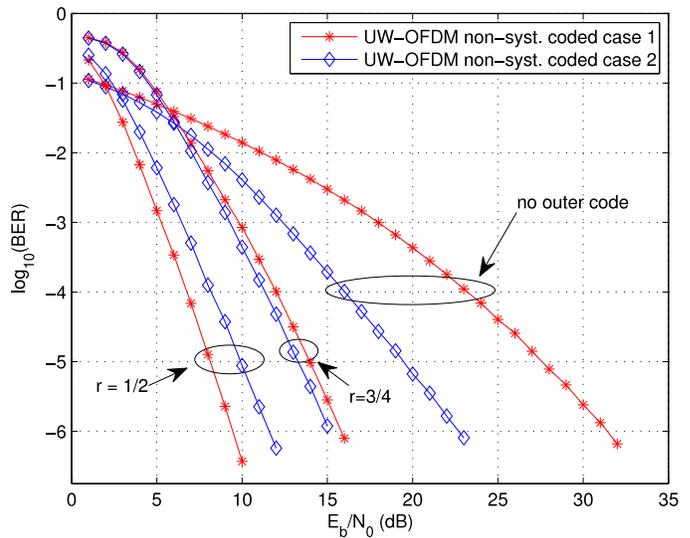}
\caption{\footnotesize{Comparison of the different non-systematic coded UW-OFDM
    systems: $\breve{\m{G}}'$ versus $\breve{\m{G}}''$ (QPSK).}}
\label{fig:BER_QPSK_5000channels_case_1_and_2}
\end{figure}
Fig. \ref{fig:BER_QPSK_5000channels_case_1_and_2} shows that $\breve{\m{G}}''$
features extremely good performance without an outer code, it significantly
outperforms $\breve{\m{G}}'$ in that case. For a coding rate of $r=\tfrac{3}{4}$,
$\breve{\m{G}}''$ still performs slightly better than $\breve{\m{G}}'$, while
for $r=\tfrac{1}{2}$, $\breve{\m{G}}'$ clearly outperforms
$\breve{\m{G}}''$. The coding gain achieved by a strong outer code in a
frequency selective channel is high for $\breve{\m{G}}'$ as it might be
expected for a system rather related to classical OFDM, while it is comparably
low for $\breve{\m{G}}''$ with its particular dispersive preprocessing properties making the transceiver rather behave like a single carrier system.

\subsection{BER Simulation Results for Imperfect Channel Knowledge}

Up till now we have presented performance results assuming perfect channel
knowledge at the receiver. In this subsection we investigate the effect of
channel estimation errors on the BER performance. Since in UW-OFDM the channel
$\mf{H}$ is incorporated into the receiver processing quite in a different way
as in CP-OFDM, it is not immediately obvious whether channel estimation errors
will degrade the systems' BER performance in the same scale. While in CP-OFDM
the data estimator is simply given by $\m{E}=\mf{H}^{-1}$, possible data
estimators for UW-OFDM are given by \eqref{equ:ZF_008} and
\eqref{equ:LMMSE_005}, respectively. $\mf{H}$ is usually replaced by an
estimated version $\widehat{\mf{H}}$, which has a degrading impact on the BER
performance. In UW-OFDM channel estimation errors have an additional impact,
namely in the processing step where the influence of the UW is subtracted from
the received symbol; cf. \eqref{equ:system_model_002}-\eqref{equ:system_model_001}. 
 
To investigate the influence of channel estimation errors we apply a standard preamble based channel estimation procedure, which we briefly describe below. We use the IEEE 802.11a preamble defined in \cite{IEEE99}. This preamble contains two identical BPSK (binary phase shift keying) modulated OFDM symbols (preceded by a guard interval) dedicated to channel estimation which we denote by $\ve{x}_p=\ve{x}_{p_1}=\ve{x}_{p_2} \in\mathbb{C}^{N\times 1}$. Note that for the downsized frequency domain version $\vef{x}_{p,d}=\m{B}^T \m{F}_N \ve{x}_p$ we have $\vef{x}_{p,d} \in \{-1,1\}^{(N_d+N_r)\times 1}$. Let $\ve{y}_{p_1}$ and $\ve{y}_{p_2}$ be the received noisy preamble symbols, and let $\ve{\tilde{\bar{y}}}_{p,d}=\frac{1}{2}\m{B}^T \m{F}_N (\ve{y}_{p_1}+\ve{y}_{p_2})$. Then a first course unbiased estimate of the vector of channel frequency response coefficients naturally follows as 
\begin{equation}
	\widehat{\vef{h}}_1[k] = \frac{\ve{\tilde{\bar{y}}}_{p,d}[k]}{\vef{x}_{p,d}[k]} = \ve{\tilde{\bar{y}}}_{p,d}[k] \vef{x}_{p,d}[k] \label{equ:ch_est1}
\end{equation}
for $k=0,...,(N_d+N_r-1)$. The latter step is true since $\vef{x}_{p,d}[k] \in \{-1,1\}$. This course channel estimate can be significantly improved or rather noise reduced by making the usually valid assumption that the channel impulse response does not exceed the guard duration $N_u$. With the vector of channel impulse response coefficients $\ve{h}\in\mathbb{C}^{N_u\times 1}$ and its zero padded version $\ve{h}_c\in\mathbb{C}^{N\times 1}$, respectively, this assumption can be incorporated by modelling the course channel estimate as
\begin{eqnarray}
	\widehat{\vef{h}}_1 &=& \m{B}^T\m{F}_N\ve{h}_c + \vef{n} \nonumber \\ 																																							&=& \m{B}^T\m{F}_N\begin{bmatrix}\ve{h}\\ \ve{0}\end{bmatrix} + \vef{n}, \label{equ:ch_est2}
\end{eqnarray}
where $\vef{n}\in\mathbb{C}^{(N_d+N_r)\times 1}$ represents a white Gaussian (frequency domain) noise vector. By decomposing the DFT matrix as $\m{F}_N = \begin{bmatrix}\m{M}_1 & \m{M}_2\end{bmatrix}$ with $\m{M}_1\in\mathbb{C}^{N\times N_u}$ and $\m{M}_2\in\mathbb{C}^{N\times (N-N_u)}$, \eqref{equ:ch_est2} can be rewritten as
\begin{equation}
	\widehat{\vef{h}}_1 = \m{B}^T\m{M}_1\ve{h} +\vef{n}. \label{equ:ch_est3}
\end{equation}
From the linear model in \eqref{equ:ch_est3} the MVU estimator of the channel impulse response follows to
\begin{equation}
	\widehat{\ve{h}}=\left(\m{M}_1^H\m{B}\m{B}^T\m{M}_1\right)^{-1}\m{M}_1^H\m{B}\widehat{\vef{h}}_1,
\end{equation}
cf. \cite{Kay93}. Going back to frequency domain, and again excluding the zero subcarriers from further operation, delivers the final and highly noise reduced frequency domain channel estimate 
\begin{eqnarray}
	\widehat{\vef{h}}_2 &=&\m{B}^T\m{F}_N\begin{bmatrix}\widehat{\ve{h}} \\ \ve{0}\end{bmatrix} \nonumber \\													&=&\underbrace{\m{B}^T\m{M}_1\left(\m{M}_1^H\m{B}\m{B}^T\m{M}_1\right)^{-1}\m{M}_1^H\m{B}}_{\m{W}}\widehat{\vef{h}}_1.
		\label{equ:ch_est4}
\end{eqnarray}
Note that the smoothing matrix $\m{W}\in\mathbb{C}^{(N_d+N_r)\times(N_d+N_r)}$ does not depend on the channel, and has to be calculated only once during system design. The preamble based estimate of the channel matrix is therefore given by $\widehat{\mf{H}} = \text{diag}\{\widehat{\vef{h}}_2\}$. \smallskip

\begin{figure}[t]
\centering
\includegraphics[width=3.5in]{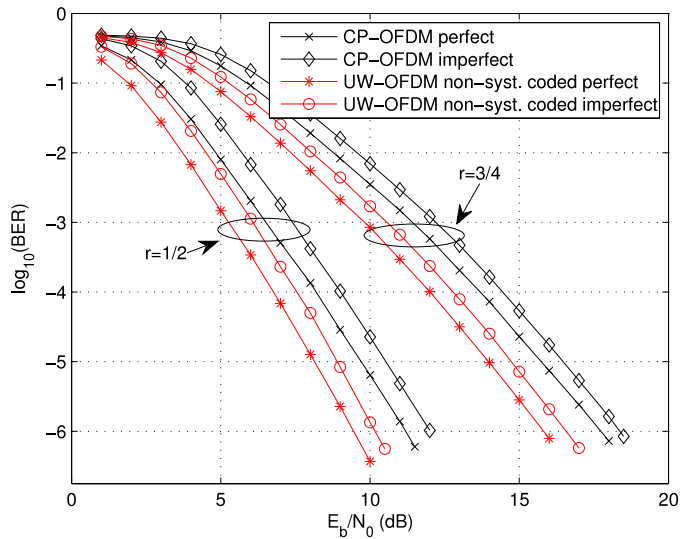}
\caption{Impact of imperfect channel estimation on the BER performance of
  CP-OFDM and non-systematic coded UW-OFDM ($\m{G}'$) in a frequency selective
environment with additional outer coding (QPSK).}
\label{fig:BER_QPSK_5000channels_imperfect_ch_estimation}
\end{figure}

Fig. \ref{fig:BER_QPSK_5000channels_imperfect_ch_estimation} compares the performance loss of CP-OFDM and non-systematic coded UW-OFDM ($\breve{\m{G}}'$) in case the described preamble based channel estimate given by \eqref{equ:ch_est4} is used instead of perfect channel knowledge. As a highly interesting result we notice that both systems degrade by about the same scale: CP-OFDM experiences a loss of 0.8dB for $r=1/2$ and 0.6dB for $r=3/4$, respectively (all results again measured at a BER of $10^{-6}$), while the performance of UW-OFDM degrades by 0.7dB for both coding rates.

\section{Conclusion}
In this work we expanded our recently introduced systematic coded UW-OFDM
concept to non-systematic coded UW-OFDM. For that we introduced optimized code
generator matrices that distribute the redundancy over all subcarriers instead
of only over a dedicated set. We derived optimization criteria to find a class
of code generator matrices that in case of AWGN conditions ensure minimum error
variances on the subcarriers after the estimation process at the
receiver. However, due to the advantageous combination of the channel coding
and dispersive preprocessing abilities of the optimized code generator matrices, non-systematic coded UW-OFDM particularly features its superior performance in frequency selective channels. We showed simulation results for selected code generator matrices in the AWGN case as well as in frequency selective environments. It turns out that non-systematic coded UW-OFDM impressively outperforms systematic coded UW-OFDM and classical CP-OFDM w.r.t. the spectral and the bit error ratio behavior. 

\section*{Acknowledgment}
\noindent
The authors want to express their deep thanks to the anonymous reviewers for many valuable comments.

\appendices

\section{Derivation of the Global Minimum of $J_\text{BLUE}$} \label{sec:appendix_a}
In Appendix \ref{sec:appendix_a} we proof that the gradient of the cost function $J_\text{BLUE}$ with respect to the vector $\ve{s}=\begin{bmatrix} s_1 & s_2 & \cdots & s_{N_d} \end{bmatrix}^T$ of singular values of $\breve{\m{G}}$ is zero if and only if all singular values are identical. And we proof that every local minimum of the cost function $J_\text{BLUE}$ is also a global minimum with $J_\text{BLUE,min} = \sigma_d^2 N_d/c$. Using the SVD in \eqref{equ:an001} we have
\begin{align}
	\mathrm{tr}\{\breve{\m{G}}^H\breve{\m{G}}\} &= \mathrm{tr}\{\m{V}\m{\Sigma}^H\m{U}^H\m{U}\m{\Sigma}\m{V}^H\} 
																							 =	\mathrm{tr}\{\m{V}\m{D}^2\m{V}^H\} \nonumber \\
																			&=   			\mathrm{tr}\{\m{D}^2\}, \label{equ:an002} \\
	(\breve{\m{G}}^H\breve{\m{G}})^{-1} &= (\m{V}\m{\Sigma}^H\m{U}^H\m{U}\m{\Sigma}\m{V}^H)^{-1} 
																							 =  (\m{V}\m{D}^2\m{V}^H)^{-1} \nonumber \\
																			&= \m{V}(\m{D}^2)^{-1}\m{V}^H. \label{equ:an003}
\end{align}
Inserting \eqref{equ:an002} and \eqref{equ:an003} into \eqref{equ:syst007} leads to the following expression for the cost function
\begin{align}
	J_\text{BLUE} &= \frac{\sigma_d^2}{cN_d}\mathrm{tr}\{\m{D}^2\}\mathrm{tr}\left\{\m{V}(\m{D}^2)^{-1}\m{V}^H\right\} 
										\nonumber \\
								&= \frac{\sigma_d^2}{cN_d}\mathrm{tr}\{\m{D}^2\}\mathrm{tr}\left\{(\m{D}^2)^{-1}\right\} \nonumber \\
								&= \frac{\sigma_d^2}{cN_d} \left(s_1^2+s_2^2+\cdots s_{N_d}^2\right)
										\left(\frac{1}{s_1^2}+\frac{1}{s_2^2}+\cdots \frac{1}{s_{N_d}^2}\right), \nonumber \\ \label{equ:an004}
\end{align}
which can now be regarded as a function of the singular values of $\breve{\m{G}}$. The gradient of $J_\text{BLUE}$ with respect to $\ve{s}$ follows to
\begin{eqnarray}
	\lefteqn{\frac{\partial J_\text{BLUE}}{\partial \ve{s}} = \frac{2\sigma_d^2}{cN_d} \times} \nonumber \\
		& & \times \begin{bmatrix}
		s_1\left(\frac{1}{s_1^2}+\cdots \frac{1}{s_{N_d}^2}\right)-s_1^{-3}\left(s_1^2+\cdots s_{N_d}^2\right)	\\
		\vdots \\
		s_{N_d}\left(\frac{1}{s_1^2}+\cdots \frac{1}{s_{N_d}^2}\right)-s_{N_d}^{-3}\left(s_1^2+\cdots	s_{N_d}^2\right)
		\end{bmatrix}. \nonumber \\
\end{eqnarray}
Setting the gradient to zero leads to the system of equations
\begin{eqnarray}
	s_1^4\left(\frac{1}{s_1^2}+\cdots \frac{1}{s_{N_d}^2}\right) &=& s_1^2+\cdots	s_{N_d}^2 \nonumber \\
	\vdots \nonumber \\
	s_{N_d}^4\left(\frac{1}{s_1^2}+\cdots \frac{1}{s_{N_d}^2}\right) &=& s_1^2+\cdots	s_{N_d}^2.
\end{eqnarray}
It is easy to see that $s_1 = s_2 = \cdots s_{N_d}:=s$ solves the system of
equations. Furthermore, by subtracting the $i^\text{th}$ equation from the
$j^\text{th}$ equation for all $i,j\in \{1,...,N_d\}$ with $i\neq j$ it
immediately follows that $s_1 = s_2 = \cdots s_{N_d}$ is in fact the only
solution to this system of equations. Consequently, every possible
candidate $\breve{\m{G}}$ for a local minimum satisfies
$\breve{\m{G}}^H\breve{\m{G}} = s^2\m{I}$ (cf. \eqref{equ:num010} and its
implications). Inserting $\breve{\m{G}}^H\breve{\m{G}} = s^2\m{I}$ into the
cost function \eqref{equ:syst007} leads to the same expression as in
\eqref{equ:syst030} that corresponds to the numerically found local
minima. Hence, every $\breve{\m{G}}$ fulfilling
$\breve{\m{G}}^H\breve{\m{G}} = s^2\m{I}$ results in the same (and minimum)
value $J_\text{BLUE,min} = \sigma_d^2 N_d/c$ which therefore constitutes the
global minimum of the cost function.

\section{Derivation of the Global Minimum of $J_\text{LMMSE}$} \label{sec:appendix_b}
In Appendix \ref{sec:appendix_b} we proof that $\partial J_\text{LMMSE}/ \partial \ve{s}=\ve{0}$ if and only if all singular values are identical. And we proof that every local minimum of the cost function $J_\text{LMMSE}$ is also a global minimum with $J_\text{LMMSE,min} = \sigma_d^2 N_d/(c+1)$.  Inserting \eqref{equ:an002} and \eqref{equ:an003} into \eqref{equ:syst035} leads to the following expression for the cost function
\begin{equation}
	J_\text{LMMSE} = \sigma_d^2 \mathrm{tr}\left\{\left(
									 \frac{c N_d}{\mathrm{tr}\{\m{D}^2\}} \m{V}\m{D}^2\m{V}^H+\m{I}\right)^{-1}\right\},  \label{equ:app001}
\end{equation}
which can be regarded as a function of the singular values of $\breve{\m{G}}$. Applying the matrix inversion lemma we immediately obtain
\begin{align}
	J_\text{LMMSE} &= \sigma_d^2 \mathrm{tr}\left\{\m{I}-\m{V}\left(\m{V}^H\m{V}+\frac{\mathrm{tr}\{\m{D}^2\}}{c N_d}
										(\m{D}^2)^{-1}\right)^{-1}\m{V}^H  \right\} \nonumber \\
								 &= \sigma_d^2 \left(N_d -\mathrm{tr}\left\{\left(\m{I}+\frac{\mathrm{tr}\{\m{D}^2\}}{c N_d}
										(\m{D}^2)^{-1}\right)^{-1}  \right\}\right). 	\label{equ:app002}
\end{align}
With this step we achieved, that every matrix to be inverted in \eqref{equ:app002} has a diagonal structure. Having in mind that $\m{D}^2$ is a diagonal matrix with the squared singular values of $\breve{\m{G}}$ on its main diagonal, and after some rearrangements we obtain
\begin{equation}
	J_\text{LMMSE} = \sigma_d^2 N_d-\sigma_d^2 c N_d \sum_{i=1}^{N_d} {\frac{s_i^2}{c N_d s_i^2+\mathrm{tr}\{\m{D}^2\}}}. \label{equ:app008}
\end{equation}
The partial derivation of the cost function $J_\text{LMMSE}$ with respect to the $j^{\text{th}}$ singular value follows to
\begin{eqnarray}
	\lefteqn{\frac{\partial J_\text{LMMSE}}{\partial s_j} = -2\sigma_d^2 c N_d s_j \times} \nonumber \\
			& & \times \sum_{\{i:i\neq j\}} \left[{\frac{s_i^2}{(c N_d s_j^2+\mathrm{tr}\{\m{D}^2\})^2}} 
					- {\frac{s_i^2}{(c N_d s_i^2+\mathrm{tr}\{\m{D}^2\})^2}}\right]. \nonumber \\
\end{eqnarray}
It is easy to see that $\partial J_\text{LMMSE}/\partial s_j=0$ is fulfilled if
$s_i=s_j$ for all $i\in \{1,...,N_d\}$ with $i\neq j$. In fact $s_1=\cdots =
s_{N_d}=s$ is the only solution to $\partial J_\text{LMMSE}/\partial
\ve{s}=\ve{0}$. This can be proved by subtracting the equations resulting from
$\partial J_\text{LMMSE}/\partial s_i=0$ and $\partial J_\text{LMMSE}/\partial
s_j=0$ for all $i\neq j$, which is not difficult but a kind of exhausting. The
remaining argumentation coincides with the one for the BLUE in Appendix
\ref{sec:appendix_a}. However, the expression for the global minimum
$J_\text{LMMSE,min} = \sigma_d^2 N_d/(c+1)$ which is obtained by inserting
$\breve{\m{G}}^H\breve{\m{G}} = s^2\m{I}$ into \eqref{equ:syst035} differs from $J_\text{BLUE,min}$.

\begin{IEEEbiography}[{\includegraphics[width=1in,height=1.25in,clip,keepaspectratio]{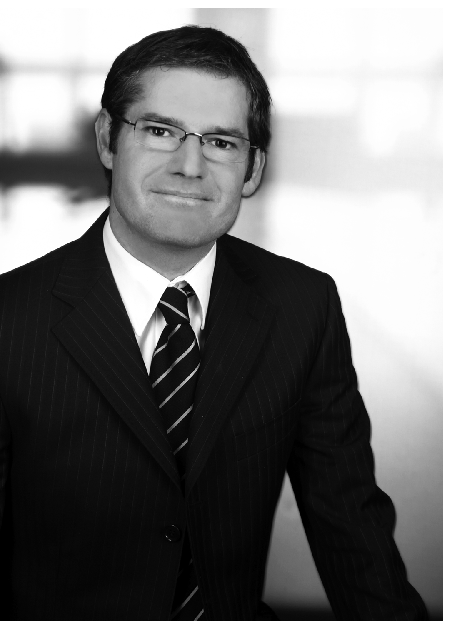}}]{Mario Huemer}
was born in Wels, Austria in 1970. He received the Dipl.-Ing. degree in 
mechatronics and the Dr.techn. (Ph.D.) degree from the Johannes Kepler University of Linz, 
Austria, in 1996 and 1999, respectively. From 1997 to 2000, he was a scientific assistant 
at the Institute for Communications and Information Engineering at the University of Linz, 
Austria. From 2000 to 2002, he was with Infineon Technologies Austria, research and development 
center for wireless products. From 2002-2004 he was a Professor for Communications and 
Information Engineering at the University of Applied Sciences of Upper Austria, from 2004-2007 
he was Associate Professor for Electronics Engineering at the University of Erlangen-Nuremberg, 
Germany. In 2007, he has moved to Klagenfurt, Austria, to establish the Chair of Embedded
Systems and Signal Processing at Klagenfurt University as a Full Professor. Mario Huemer has been engaged in research and development on WLAN, wireless cellular, and wireless 
positioning systems, and on highly integrated baseband and RFICs for mobile
devices. He published more than 130 papers in international journals or
conference records. His current research interests are focused on 
signal processing algorithms, architectures and applications.
 
Mario Huemer is a member of the IEEE Signal Processing Society, IEEE Communications Society 
and the IEEE Circuits and Systems Society. He is also a member of the European 
Association for Signal Processing (EURASIP), the German Information Technology Society (ITG) 
in the Association for Electrical, Electronic and Information Technologies (VDE) and the 
Austrian Electrotechnical Association (ÖVE).
\end{IEEEbiography}

\begin{IEEEbiography}[{\includegraphics[width=1in,height=1.25in,clip,keepaspectratio]{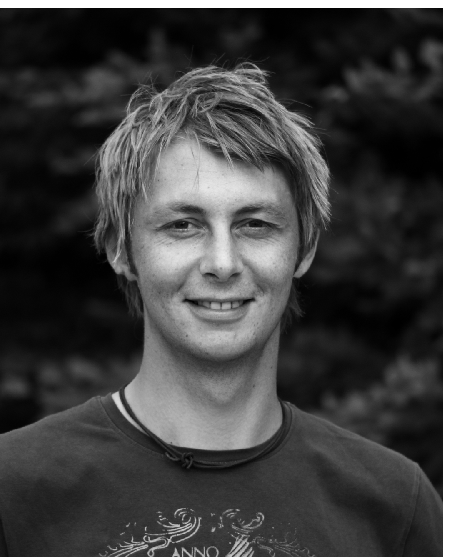}}]{Christian Hofbauer}
was born in 1982 in St. Peter am Wimberg, Austria. Between 2002 and
2006, he studied hardware/software systems engineering at the University
of Applied Sciences of Upper Austria and received the Dipl.-Ing. (FH)
degree. He conducted research for his thesis "High level refinement and
implementation cost estimation for Matlab/Simulink models, targeting
handheld Software Defined Radio architectures" at the Research Institute
IMEC, Belgium. Since September 2007, he has been a member of the
Embedded Systems and Signal Processing Group, Alpen-Adria-Universit\"at
Klagenfurt, Austria. Currently, his research focuses on the
investigation of the UW-OFDM concept.
\end{IEEEbiography}

\begin{IEEEbiography}[{\includegraphics[width=1in,height=1.25in,clip,keepaspectratio]{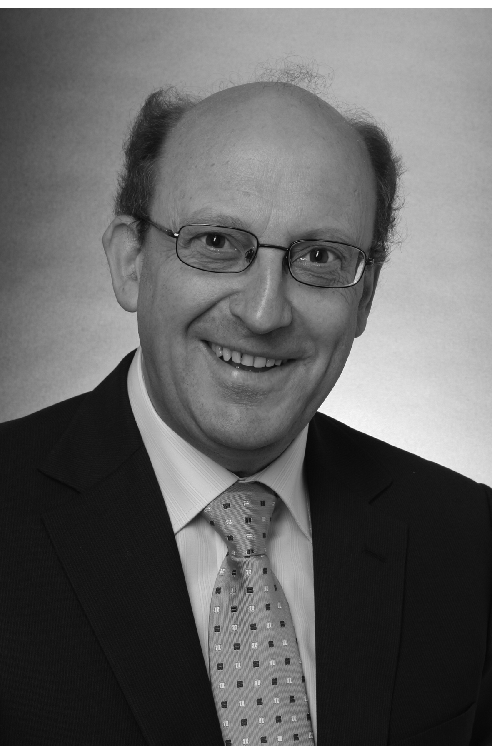}}]{Johannes
    B. Huber}
received the Dipl.-Ing. degree in electrical engineering from Munich Technical University 1977. From 1977 to 1991 he has been a research assistant and chief engineer at the Federal Armed Forces University, Munich,, from which he received the Dr.-Ing. degree with a thesis on coding for channels with memory in 1982 and the Dr. habil. degree with a thesis on trellis coded modulation in 1991. In 1991, he joined the IBM Research Laboratory, Zurich, Switzerland. 
 
Since autumn 1991, he is full Professor at the Universiy of Erlangen-Nuremberg, Germany. His research interests are information and coding theory, modulation schemes, algorithms for signal detection and equalization for channels with severe intersymbol interference, signalling, detection and equalization for multiple-input multiple-output (MIMO) channels and concatenated coding together with iterative decoding.
 
Johannes Huber is Fellow of the IEEE and he had been elected several times for a member of the board of governors of the IEEE Information Theory Society. He is an ordinary member of the Bavarian Academy of Science and an Corresponding Fellow of the Royal Society of Edinburgh.
 
From 1994 to 2004, Johannes Huber was a member of the Editorial Board of the "International Journal on Electronics and Communications (AEÜ)" and from1997 to 2002 Editor-in-Chief of this journal. From 1996 to 1999 he was an Associate Editor of the IEEE Transactions on Communications. He has also been general chairman and chairman of the program committees of several international conferences on information theory and coding.
 
Johannes Huber is author or co-author of textbooks on trellis coding and information combining. He published about 240 papers in international journals or conference records. Papers authored or co-authored by him were awarded by the Best Paper Award of the German Society of Information Technology (ITG) in 1988, 2000, and 2006. He also received the Vodafone award for innovations in mobile communications 2004.
\end{IEEEbiography}


\begin{thebibliography}{1}



\bibitem{Huemer10_1}
M. Huemer, C. Hofbauer, J.B. Huber, ``The Potential of Unique Words in OFDM,'' in the \emph{Proceedings of the 15th International OFDM-Workshop}, Hamburg, Germany, pp. 140-144, September 2010.

\bibitem{Onic10_1}
A. Onic, M. Huemer, ``Direct versus Two-Step Approach for Unique Word
Generation in UW-OFDM,'' in the \emph{Proceedings of the 15th International OFDM-Workshop}, Hamburg, Germany, pp.145-149, September 2010.

\bibitem{VanNee00}
R. van Nee, R. Prasad, \emph{OFDM for Wireless Multimedia Communications}, Artech House Publishers, Boston, 2000.

\bibitem{Tang07}
S. Tang, F. Yang, K. Peng, C. Pan, K. Gong, Z. Yang, ``Iterative channel estimation for block transmission with known symbol padding - a new look at TDS-OFDM,'' in the \emph{Proceedings of the IEEE Global Telecommunications Conference (GLOBECOM 2007)}, pp. 4269-4273, Nov. 2007.

\bibitem{Welden08}
D. Van Welden, H. Steendam, M. Moeneclaey, ``Iterative DA/DD channel estimation for KSP-OFDM,'' in the \emph{Proceedings of the IEEE International Conference on CommunicationsICC 2008)}, pp. 693-697, May 2008.

\bibitem{Lin11}
Y. P. Lin, S. M. Phoong, P. P. Vaidyanathan, \emph{Filter Bank Transceivers for OFDM and DMT Systems}, Cambridge University Press, Cambridge 2011.

\bibitem{Blahut03}
R.E. Blahut, \emph{Algebraic Codes for Data Transmission}, Cambridge University Press, New York, 2003.

\bibitem{Hofbauer10_03}
C. Hofbauer, M. Huemer, J. B. Huber, ``Coded OFDM by Unique Word Prefix,'' in the \emph{Proceedings of the IEEE International Conference on Communication Systems (IEEE ICCS' 2010)}, Singapore, 5 pages, November 2010. 

\bibitem{Huemer11_1}
M. Huemer, A. Onic, C. Hofbauer, ``Classical and Bayesian Linear Data
Estimators for Unique Word OFDM,'' \emph{IEEE Transactions on Signal
  Processing}, vol. 59, no. 12, pp. 6073-6085, Dec. 2011.

\bibitem{Sari94a}
H. Sari, G. Karam, I. Jeanclaude, ``An Analysis of Orthogonal Frequency-Division Multiplexing for Mobile Radio Applications,'' in the \emph{Proceedings of the IEEE Vehicular Technology Conference (VTC '94)}, Stockholm, Sweden, pages 1635-1639, June 1994.

\bibitem{Sari94b}
H. Sari, G. Karam, I. Jeanclaude, ``Frequency-Domain Equalization of Mobile Radio and Terrestrial Broadcast Channels,'' in the \emph{Proceedings of the IEEE International Conference on Global Communications (GLOBECOM '94)}, San Francisco, USA, pages 1-5, 1994.

\bibitem{Czy97a}
A. Czylwik, ``Comparison between Adaptive OFDM and Single Carrier Modulation with Frequency Domain Equalization,'' in the \emph{Proceedings of the IEEE Vehicular Technology Conference (VTC '97)}, Phoenix, USA, pages 865-869, May 1997.

\bibitem{Kad97a}
G. Kadel, ``Diversity and Equalization in the Frequency Domain - a Robust and Flexible Receiver Technology for Broadband Mobile Communication Systems,'' in the \emph{Proceedings of the IEEE Vehicular Technology Conference (VTC '97)}, Phoenix, USA, pages 894-898, May 1997.

\bibitem{Clark98}
M. V. Clark, ``Adaptive frequency-domain equalization and diversity combining for broadband wireless communications,'' in the \emph{IEEE Journal on Selected Areas in Communications}, Vol. 16, No. 8, pages 1385-1395, October 1998.

\bibitem{Huemer99a}
M. Huemer, L. Reindl, A. Springer, R. Weigel, ``Implementation Aspects on Single Carrier Transmission with Frequency Domain Equalization,'' in the \emph{Proceedings of the 4th International OFDM-Workshop '99}, Hamburg, Germany, pages 18.1-18.4, September 1999.

\bibitem{Huemer99b}
M. Huemer, ``Frequenzbereichsentzerrung f{\"u}r hochratige Eintr{\"a}ger-{\"U}bertragungssysteme in Umgebungen mit ausgepr{\"a}gter Mehrwegeausbreitung,'' Dissertation, Institute for Communications and Information Engineering, University of Linz, Austria, 1999 (in German).

\bibitem{Imec00}
L.~Deneire, B.~Gyselinckx, M.~Engels, ``Training Sequence vs. Cyclic Prefix: A New Look on Single Carrier Communication,'' in the \emph{Proceedings of the IEEE International Conference on Global Communications (GLOBECOM '2000)}, pages 1056-1060, November 2000.

\bibitem{Cendrillon01}
R. Cendrillon, M. Moonen, ``Efficient equalizers for single and multi-carrier environments with known symbol padding,'' in the \emph{Proceedings of the IEEE International Symposium on Signal Processing and its Applications (ISSPA 2001)}, August 2001, Pages 607-610.

\bibitem{Witschnig02a}
H. Witschnig, T. Mayer, A. Springer, A. Koppler, L. Maurer, M. Huemer,R. Weigel, ``A Different Look on Cyclic Prefix for SC/FDE,'' in the \emph{Proceedings of the 13th IEEE International Symposium on Personal, Indoor and Mobile Radio Communications (PIMRC 2002)}, Lisbon, Portugal, pages 824-828, September 2002. 

\bibitem{Witschnig02b}
H. Witschnig, T. Mayer, A. Springer, L. Maurer, M. Huemer, R. Weigel, ``The Advantages of a Known Sequence versus Cyclic Prefix in an SC/FDE System,'' in the \emph{Proceedings of the 5th International Symposium on Wireless Personal Multimedia Communications (WPMC' 2002)}, Honolulu, Hawaii, pages 1328-1332, October 2002.

\bibitem{Huemer03b}
M. Huemer, A. Koppler, L. Reindl, R. Weigel, ``A Review of Cyclically Extended Single Carrier Transmission with Frequency Domain Equalization for Broadband Wireless Transmission,'' in the \emph{European Transactions on Telecommunications (ETT)}, Vol. 14, No. 4, pages 329-341, July/August 2003.

\bibitem{Huemer03a}
M. Huemer, H. Witschnig, J. Hausner, ``Unique Word Based Phase Tracking Algorithms for SC/FDE Systems,'' in the \emph{Proceedings of the IEEE International Conference on Global Communications (GLOBECOM' 2003)}, San Francisco, USA, 5 pages, December 2003.

\bibitem{Witschnig03a}
H. Witschnig, ``Frequency Domain Equalization for Broadband Wireless Communication - With Special Reference to Single Carrier Transmission Based on Known Pilot Sequences,'' Dissertation, University of Linz, Institute for Communications and Information Engineering, 2004.

\bibitem{Reinhardt05}
S. Reinhardt, T. Buzid, M. Huemer, ``MIMO Extensions for SC/FDE Systems,'' in the \emph{Proceedings of the European Conference on Wireless Technology (ECWT' 2005)}, Paris, France, pp. 109-112, October 2005.

\bibitem{Huemer10_2}
M. Huemer, C. Hofbauer, J.B. Huber, ``Unique Word Prefix in SC/FDE and OFDM: A Comparison,'' in the \emph{Proceedings of the IEEE GLOBECOM 2010 Workshop on Broadband Single Carrier and Frequency Domain Communications (BSCFDC 2010)}, Miami, USA, pp. 1321-1326, December 2010. 

\bibitem{Muck06}
M. Muck, M. de Courville, P. Duhamel, ``A pseudorandom postfix OFDM modulator - semi-blind channel estimation and equalization,'' in the \emph{IEEE Transactions on Signal Processing}, Vol. 54, Issue 3, pp 1005-1017, March 2006.

\bibitem{Jingyi02}
L. Jingyi, P. Joo, J. Ro, ``The effect of filling Unique Words to guard interval for OFDM,'' Document IEEE C802.16a-02/87, IEEE 802.16 Broadband Wireless Access Working Group, September 2002.

\bibitem{Hofbauer10_2}
C. Hofbauer, M. Huemer, J.B. Huber, ``On the Impact of Redundant Subcarrier Energy Optimization in UW-OFDM,'' in the \emph{Proceedings of the 4th International Conference on Signal Processing and Communication Systems (ICSPCS 2010)}, Gold Coast, Australia, 6 pages, December 2010.
  
\bibitem{Huemer10_3}
M. Huemer, C. Hofbauer, J.B. Huber, ``Complex Number RS Coded OFDM with Systematic Noise in the Guard Interval,'' in the \emph{Proceedings of the 44th ASILOMAR Conference on Signals, Systems and Computers}, Pacific Grove, USA, pp. 1023-1027, November 2010.

\bibitem{IEEE99}
IEEE Std 802.11a-1999, Part 11: Wireless LAN Medium Access Control (MAC) and Physical Layer (PHY) specifications: High-Speed Physical Layer in the 5 GHz Band, 1999.

\bibitem{Kay93}
S. Kay, \emph{Fundamentals of Statistical Signal Processing: Estimation Theory}, Prentice Hall, Rhode Island 1993.

\bibitem{Fak97}
J. Fakatselis, Criteria for 2.4 GHz PHY Comparison of Modulation Methods. Document IEEE 1997; P802.11-97/157r1.


\end{thebibliography}
\end{document}